\documentclass[%
 reprint,
nofootinbib,
 amsmath,amssymb,
 aps,
]{revtex4-2}

\usepackage{graphicx}
\usepackage{dcolumn}
\usepackage{bm}
\usepackage{hyperref}
\usepackage{xcolor}
\usepackage{caption}
\usepackage{subcaption}
\usepackage{float}
\usepackage{tikz}
\usepackage{diagbox}
\usepackage{amsmath}
\usepackage{amssymb}
\usepackage{tabularx}

\begin{document}

\preprint{APS/123-QED}

\title{Electron Capture and Bound-State $\beta$ Decays in Ions and Plasma}

\author{Bharat Mishra}
\email{mishra@lns.infn.it}
\author{Angelo Pidatella}%
 \email{pidatella@lns.infn.it}
 \author{David Mascali}
\affiliation{%
 Istituto Nazionale di Fisica Nucleare - Laboratori Nazionali del Sud (INFN - LNS), Catania, Italy
}
\author{Simone Taioli}%
 \affiliation{%
 European Centre for Theoretical Studies in Nuclear Physics and Related Areas - Bruno Kessler Foundation (ECT* - FBK), Trento, Italy\\
 Trento Institute for Fundamental Physics and Applications (INFN - TIFPA), Trento, Italy
}
\author{Stefano Simonucci}%
 \affiliation{%
 School of Science and Technology, University of Camerino, Camerino, Italy\\
Istituto Nazionale di Fisica Nucleare - Sezione di Perugia, Perugia, Italy
}%


\date{\today}

\begin{abstract}
We present a new theory describing the variation of electron capture and bound-state $\beta$-decays in atomic ions and (non) local thermodynamic equilibrium ((N)LTE) plasmas. We adopt the Takahashi-Yokoi nuclear model with added corrections to first calculate the decay rate for each atomic configuration of the isotope, and then evaluate the in-plasma decay rate by combining them with the charge state distribution (CSD) consistent with plasma density and temperature. Our approach expands the thermodynamic parameter space in which in-plasma $\beta$-decays can be studied, opening the possibility to validate the model in low-density laboratory magnetoplasmas before application to stellar nucleosynthesis. The model is explained using $^{7}$Be, and then applied to higher mass isotopes such as $^{140}$Pr$^{0+,57+,58+}$, $^{142}$Pm$^{0+,59+,60+}$ and $^{163}$Dy$^{66+}$. Our model is therefore amenable to isotopes in a wide range of masses, in both single charge state or in a plasma-generated CSD. 
\end{abstract}

\maketitle


\section{Introduction}
\label{Intro}

Modification of $\beta$-decay rates $\lambda$ due to changes in the atomic environment of radioactive nuclei has been known for decades~\cite{decayrev1972}. Early experiments on the subject observed the effect to a small extent by changing the chemistry of the material containing the decaying nuclei~\cite{Bouchez1949,Leininger1949,Ray1999,Norman2001,Ohtsuki2007}. The effect was attributed to perturbation of the electron density on the nuclear surface which affected electron capture (EC) rates~\cite{DaudelBDecay1947,Morisato2008,Lee2008}, in line with the the hypotheses of Daudel and Segre~\cite{Daudel1947,Segre1947}. A much stronger effect of the atomic environment on $\lambda$ was observed in highly charged ions (HCIs) circulating in storage rings where the removal of bound electrons modified existing EC channels or activated hitherto prohibited bound state $\beta$-decay (BSBD) channels~\cite{Litvinov2023}. As a result, stable isotopes like $^{163}$Dy became suddenly unstable~\cite{Jung1992} while already radioactive isotopes like $^{187}$Re underwent nine orders of magnitude reduction in $t_{1/2}$ when stripped of all electrons~\cite{Bosch1996}.

Stellar nucleosynthesis models developed for calculating $s$-process branching of heavy nuclei are sensitive to both neutron capture cross sections $\sigma(n,\gamma)$ and $\lambda$~\cite{Arcones2022}. Uncertainties in either quantity can strongly affect the balance between the competing processes, leading to mismatch between observed and predicted elemental abundances~\cite{Palmerini2021,Busso2022}. While there are experimental facilities already operational for measuring and/or updating relevant $\sigma(n,\gamma)$ to high precision~\cite{Lisowski1990,nTOF2013}, nucleosynthesis models still rely on decay half-lives $t_{1/2}$ measured in neutral atoms which may be significantly different from their stellar plasma counterparts.

The interior of stars is a naturally occurring dense plasma which contains isotopes in a various charge states and excitation levels - hereafter referred to as charge state distribution (CSD) and level population distribution (LPD) - surrounded by an energetic electron cloud. It can thus be expected that decay rates will vary strongly, depending on the properties of the plasma~\cite{Palmerini2021,Busso2022}. The first comprehensive theories on $\beta$-decay in stellar plasma were proposed by Bahcall~\cite{Bahcall1961,Bahcall1962,BahcallEC1964}, and covered continuum decays and captures in great detail. A full systematics-based analysis of all in-plasma $\beta$-decay processes was made by Takahashi and Yokoi~\cite{TakahashiYokoi1983} (hereafter TY83) who put forward elegant expressions for calculating the lepton phase volume associated with each decay channel and showed how the stellar plasma may cause enhancement/suppression of some transitions. Their formulations were also used to predict BSBD rates in fully-ionised nuclei~\cite{Gupta2019,Liu2021,Gupta2023}, reproducing later observations from storage rings in the process.

In order to study plasma-induced changes in $\beta$-decay rates experimentally, a new facility named PANDORA (Plasmas for Astrophysics, Nuclear Decay Observations and Radiation for Archaeometry) is under realisation at INFN-LNS in Catania, Italy~\cite{Mascali2022}. The facility aims to use an electron cyclotron resonance (ECR) ion source as a compact magnetoplasma to measure in-plasma $\beta$-decay rates and validate the TY83 model. To this effect, the latter must be modified because unlike stellar plasmas for which TY83 was originally designed, laboratory magnetoplasmas and other astrophysical plasmas are spatially non-uniform, orders of magnitude lower in density and in non local thermodynamic equilibrium (NLTE). We generalise the TY83 model to any kind of plasma by splitting it into two components:
\begin{itemize}
    \item An \textit{atomic component}, which calculates the decay rate of the radio-isotope as a function of its charge state \textit{i} and excitation level \textit{j} - hereafter referred to as the \textit{configuration-dependent decay rate} $\lambda^{*}(ij)$
    \item A \textit{plasma component}, which calculates the CSD and LPD associated with the radio-isotope as a function of plasma density and temperature - hereafter denoted by the \textit{probability factor} $p_{ij}$
\end{itemize}
The in-plasma decay rate is then obtained by combining $p_{ij}$ with $\lambda^{*}(ij)$. In this way, it becomes possible to analyse the fundamental coupling of atomic and nuclear properties independent of the plasma, and then incorporate the effects of the latter through the $p_{ij}$ factors. In this work, we demonstrate our method by studying the variation of orbital EC decay of $^{7}$Be in a generic plasma. The isotope is one of the first physics cases to be investigated in PANDORA due to its short $t_{1/2}$ which facilitates measurability~\cite{Naselli2025}, and also due to its fundamental importance in nuclear astrophysics~\cite{Adelberger2011,Iben1967}. The formalism is identical for BSBD and has been thoroughly validated with other isotopes-  $^{140}$Pr, $^{142}$Pm and $^{163}$Dy - which have been already investigated in storage rings. 

\section{Methodology}
\label{Method}

\begin{figure}
    \centering
    \includegraphics[width=\linewidth]{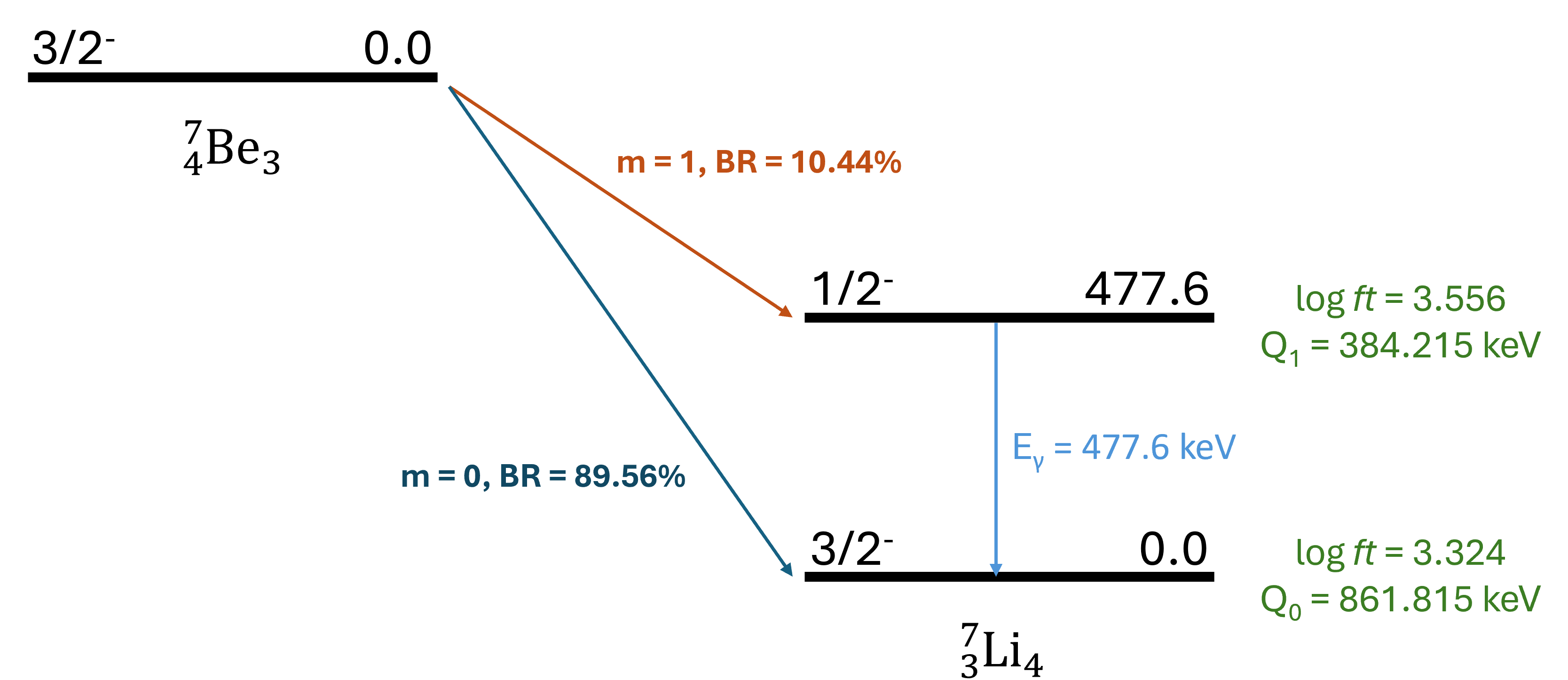}
    \caption{$^{7}$Be$\rightarrow^{7}$Li EC decay scheme, taken from ENSDF~\cite{Tilley2002}.}
    \label{Fig1}
\end{figure}

$^{7}$Be decays into $^{7}$Li through the electron capture process
\begin{equation}
    \label{Eq1}
    ^{7}\mathrm{Be}+e^{-}\rightarrow^{7}\mathrm{Li}+\nu_{e}
\end{equation}
where the captured $e^{-}$ may belong to either the continuum or an atomic orbital. In the case of the former, the decay is classified as \textit{continuum capture}, while the latter is termed as \textit{orbital capture}. The decay scheme is shown in Fig.~\ref{Fig1} and the $Q$-values and branching ratio (BR) of the two channels $m=0$ and $m=1$ are indicated. Both are allowed transitions, with the former being mixed Fermi and Gamow-Teller and the latter being pure Gamow-Teller. They are characterised by comparative half-lives $\log{(ft)_{0}}=3.324$ and $\log{(ft)_{1}}=3.556$, respectively~\cite{Tilley2002} which are simply a product of the lepton phase volume $f$, and the half-life $t_{1/2}$, and depend only on the nuclear matrix element fixed by the levels of parent and daughter nuclei. 

The lepton phase volume $f$ describes the phase space accessible to the electron and neutrino during the decay. It varies with the atomic configuration $(ij)$ of the radio-isotope ion and decay channel $m$.   According to the formulation of TY83, the configuration-dependent phase volumes $f_{m}^{*}(ij)$ of orbital EC and BSBD can both be calculated as
\begin{equation}
    \label{Eq2}
    f_{m}^{*}(ij)=\sum_{x}\sigma_{x}\frac{\pi}{2}[g_{x}\,\mathrm{or}\,f_{x}]^{2}(Q^{\prime}_{m}(ij,x))^{2}S_{(m)x}
\end{equation}
where $x$ represents the atomic orbital from where the electron is captured (in case of EC) or into which it is emitted (in case of BSBD), $Q^{\prime}_{m}(ij,x)=Q_{m}(ij,x)/m_{e}c^{2}$ with $Q_{m}(ij)$ being the decay $Q$-value which varies with the ion configuration and $x$ and $m_{e}c^{2}$ being the electron rest mass energy, $[g_{x}\,\mathrm{or}\,f_{x}]^{2}$ is the larger of electron radial wavefunction (ERWF) evaluated on the nuclear radius, and $S_{(m)x}$ is the shape factor. The quantity $\sigma_{x}$ represents the occupancy of the orbital $x$. In our model, we account for the contribution of $K$- and $L$-shell electrons alone, the corresponding orbitals for which respectively are $x=1s_{1/2}$ and $x=2s_{1/2}$.

Using Eq.~\ref{Eq2}, the total configuration-dependent EC decay rate (summed over all channels) can be calculated as
\begin{equation}
    \label{Eq3}
    \lambda^{*}(ij)=\ln{2}\bigg(\frac{f^{*}_{0}(ij)}{(ft)_{0}}+\frac{f^{*}_{1}(ij)}{(ft)_{1}}\bigg)
\end{equation}
where $(ft)_{0}$, $(ft)_{1}$ are the aforementioned $ft$ values.

\subsection{Correction Factors}
\label{Corr}

The expression of the phase volume in Eq.~\ref{Eq2}, as pertaining to EC, is equivalent to the formalism by Behrens and B\"{u}hring~\cite{Behrens1982}. Over the years, their theory has been improved through the introduction of correction factors and use of better atomic data, and the most complete model is currently implemented in the BetaShape code~\cite{Mougeot2019,Mougeot2018}. According to their approach, the lepton phase volume for allowed and forbidden unique transitions is given by
\begin{equation}
    \label{Eq4}
    f_{\varepsilon}=\sum_{x}n_{x}\frac{\pi}{2}\beta_{x}^{2}q_{x}^{2}C_{x}B_{x}\bigg[1+\sum_{x'}P_{x'}\bigg]
\end{equation}
where $n_{x}$ is the relative occupation number of the orbital $x$ (equivalent to $\sigma_{x}$) and $q_{x}=Q/m_{e}c^{2}$ is the neutrino momentum. The term $C_{x}$ is analogous to $S_{(m)x}$, defined in Bambynek \textit{et al}~\cite{Bambynek1977}. 
\begin{itemize}
    \item The term $\beta_{x}$ is called the Coulomb amplitude. In multi-electron atoms/ions, the overlap between the atomic orbital and the nucleus is described by a central Coulomb potential perturbed by screening from inner-shell electrons. The finite size of the nucleus also affects the overlap. The formalism in Ref.~\cite{Behrens1982} accounts for these effects and computes the ERWFs $f_{x}$, $g_{x}$ as a power-series expansion in radial coordinate $r$~\cite{Behrens1970}. The Coulomb amplitudes connect the regular solution of the Dirac equation with special solutions at $r=0$ and therefore better represent ERWFs in multi-electron systems.
    \item $B_{x}$ represents exchange and overlap corrections. The former accounts for the contribution of other orbitals to the capture from $x$ due to the identical nature of electrons, while the latter factors in the imperfect overlap between the orbitals of parent and daughter ions when the nuclear charge changes by one unit. These effects have been discussed at length by Bahcall~\cite{Bahcall1965} and Vatai~\cite{Vatai1970}, and generalised by Mougeot~\cite{Mougeot2018}. 
    \item The term in the square brackets accounts for the so-called \textit{shake-up} and \textit{shake-off} effects. The energy from the capture can excite the remaining electron in the orbital $x$ to an orbital $x'$ (shake-up) or ionise it into the continuum (shake-off). These processes enhance the capture phase volume according to their respective probability of occurrence, the details of which can be found in Refs.~\cite{Mougeot2018}.  
\end{itemize}

Eq.~\ref{Eq4} is rigorous for \textit{precision} calculations, but its application is limited to EC in neutral atoms in ground state alone. In principle, the formalism can be extended to any charge state $i$ and atomic level $j$ for EC and BSBD alike but it requires precise knowledge of the wavefunctions of the parent and daughter ions. 

Our objective is to develop a relatively simple model which can estimate EC and BSBD rates of isotopes of varying masses under different plasma/ionisation regimes for application to nuclear astrophysics. Given the larger uncertainty associated with the plasma and astrophysical parameters themselves, a certain degree of imprecision in the atomic component is tolerable. At the same time, the model must include include the essential physics to maintain a fair amount of accuracy when predicting configuration-dependent decay rates which can be tested in experiments independent of plasmas. For this reason, we continue using the TY83 formalism, but replace the ERWFs with Coulomb amplitudes in multi-electron ions. With regards to the exchange and overlap corrections, we suggest the use the polynomial expressions deduced by  Bahcall~\cite{Bahcall1963} and Vatai~\cite{Vatai1970} to calculate $B_{x}$ for light and intermediate mass isotopes in the range $Z=6-37$.   
 
In case of EC, Eq.~\ref{Eq2} is also amended to account for \textit{hyperfine interactions} between the nucleus and electron spin in H-like ions in ground state, an effect first hypothesised by Folan and Tsifrinovich~\cite{Folan1995} to predict the enhancement/suppression in capture rates from $K$-shells. The phenomenon is not expected to manifest in plasmas because the splitting energy is much smaller than plasma temperature, but it is a fundamental process affecting H-like ions that has already been experimentally observed in $^{140}$Pr$^{58+}$~\cite{Litvinov2007} and $^{142}$Pm$^{60+}$~\cite{Winckler2009} ions in storage rings. Including these corrections, the final expression of lepton phase volume used in our model is given by
\begin{equation}
\label{Eq5}
    f_{m}^{*}(ij)=\sum_{x}\sigma_{x}\frac{\pi}{2}\beta_{x}^{2}(Q^{\prime}_{m}(ij,x))^{2}S_{(m)x}B_{x,int}[\delta^{*}_{m}(x)]
\end{equation}
where $B_{x,int}$ is the aforementioned polynomial interpolation of exchange/overlap coefficients for $Z=6-37$, and $[\delta^{*}_{m}(x)]$ is the hyperfine splitting coefficient defined as 
\begin{equation}
\label{Eq6}
[\delta^{*}_{m}(x)]=
     \begin{cases}
        F(m), & \text{EC from}\,x=1s_{1/2}\,\text{in H-like ions } \\
        1, & \text{otherwise} \\
     \end{cases}
\end{equation} 
Here $F(m)$ depends on the spin of the parent and daughter nuclear levels involved in the transition channel $m$, as described in Ref.~\cite{Patyk2008}. A detailed explanation of the effect and its implementation in our model can be found in Appendix~\ref{HyperSplit}.

\subsection{Q-value}
\label{Energy}

The configuration-dependent $Q_{m}(ij,x)$ is usually written as 
\begin{equation}
    \label{Eq8}
    Q_{m}(ij,x)=Q_{m}+[B^{*}_{^{7}Be^{i}}-B^{*}_{^{7}Li^{i'}}]+[e^{*}_{^{7}Be^{i,j}}-e^{*}_{^{7}Li^{i',j'}}]
\end{equation}
where $Q_{m}=Q_{0}$ or $Q_{1}$, $B^{*}_{^{7}Be^{i}}-B^{*}_{^{7}Li^{i'}}$ is difference in electronic binding energies when the atoms are in the charge states $i$ and $i'$ respectively, and $e^{*}_{^{7}Be^{i,j}}-e^{*}_{^{7}Li^{i',j'}}$ is the difference in total energy of the system in excited states $j$ and $j'$. While the terms in Eq.~\ref{Eq8} correspond to $^{7}$Be, the expression is identical for all radioactive isotopes in ionised states. This formalism is an extended version of the one used by Gupta and Chang~\cite{Gupta2019,Liu2021,Gupta2023}, applicable to any $ij$-configuration, and not just fully-stripped atoms. $Q_{m}(ij,x)$ varies with both $ij$ and $x$ because the daughter configuration $i^{\prime}j^{\prime}$ depends on the parent ion \textit{and} the orbital $x$ from which the capture occurs. 

Eq.~\ref{Eq8} can be more succinctly expressed as   
\begin{equation}
    \label{Eq9}
    Q_{m}(ij,x)=Q_{m}+\Delta\epsilon^{ij,x}
\end{equation}
where the $Q$-value of the decay is simply expressed as the sum of difference in nuclear energy levels ($Q_{m}$) and atomic energy levels ($\Delta\epsilon^{ij,x}$). The latter represents the difference in the energy of the $ij$-configuration of the parent ion and $i^{\prime}j^{\prime}$-configuration of the daughter ion resulting from the capture of electron in the $x$ orbital. This term is calculated as 
\begin{equation}
    \label{Eq10}
    \Delta\epsilon^{ij,x}=\epsilon^{*}_{^{7}Be^{i,j}}-\epsilon^{*}_{^{7}Li^{i^{\prime},j^{\prime}}}
\end{equation}
where $\epsilon^{*}_{^{7}Be^{i,j}}$ and $\epsilon^{*}_{^{7}Li^{i',j'}}$ represent the total energy of their respective configurations, obtained by summing the ionisation potentials of all the preceding charge states $i/i'=0,1,...,i-1$ and the energy of the level $j/j'$ relative to the ground state. This formalism is equivalent to that of Eq.~\ref{Eq8}, with the exception that the binding energies are replaced by ionisation potentials.

We used the atomic data from the population kinetics code FLYCHK~\cite{FLYCHK2005} to calculate $\Delta\epsilon^{ij,x}$. A detailed description of the methodology used and nomenclature, spectroscopic notation, and electronic configuration of FLYCHK levels can be found in Appendix~\ref{FLYCHK}. 

Fig.~\ref{Fig3} shows the variation in $\Delta\epsilon^{ij,x}$ for various configurations of $^{7}$Be. The X-axes in the plots only list a few level names for brevity, but a complete list of all levels considered can be found in Tab.~\ref{TabEC1}-\ref{TabEC4} in Appendix~\ref{FLYCHK}. The plots demonstrate the energetics of $1s_{1/2}$ and $2s_{1/2}$-EC alone because the remaining orbitals contribute much lower to the decay. Since $Q_{0}=861.815\,\mathrm{keV}$, none of the configurations prohibit the decay since their Q-variations are on the $\mathrm{eV}$ scale, but there are fluctuations nonetheless. The situation may be different in heavier isotopes with small $Q_{m}$ where variations in $\Delta\epsilon^{ij,x}$ can open (or close) existing decay channels. 

It can be easily observed that $K$-captures are accompanied by lower decay energies ($\Delta\epsilon^{ij,1s_{1/2}}<0$) because the resultant $^{7}$Li ions are always formed in $K$-shell vacant configurations which are autoionising, unbound states lying above the ionisation threhsold. The slight rise in $\Delta\epsilon^{ij,1s_{1/2}}$ in excited levels compared to the ionic ground level is due to the higher energy of excited $^{7}$Be$^{i+}$ ions. In contrast, $\Delta\epsilon^{ij,2s_{1/2}}$ are always positive (and occasionally relatively large) because $L$-shell vacant levels in Be are \textit{not} inner-hole levels and therefore the energies of such configurations are much lower than the parent ions. It is interesting to note that while \textit{all} configurations in $^{7}$Be$^{0-2+}$ can lead to $K$-captures, only certain levels in these ions can lead to $L$-captures. In H-like Be ions, the ground and first excited levels can \textit{only} lead to $K$- and $L$-captures, respectively. In general, $\Delta\epsilon^{ij,1s_{1/2}}$ remains around $-50\,\mathrm{eV}$ for all charge states.

\begin{figure*}
\centering
    \begin{subfigure}{0.49\textwidth}
      \includegraphics[width=\textwidth]{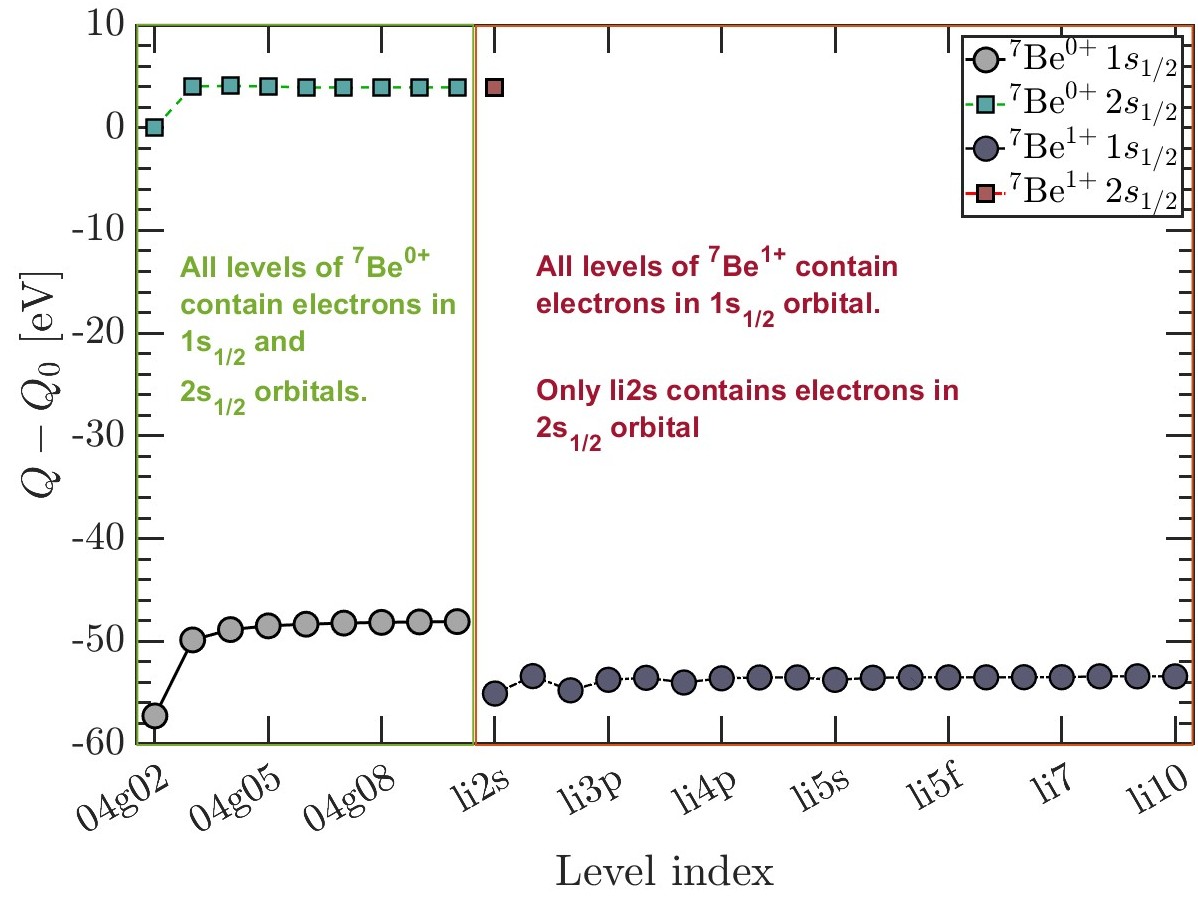}
      \caption{}
    \end{subfigure}
    \hfill
    \begin{subfigure}{0.49\textwidth}
      \includegraphics[width=\textwidth]{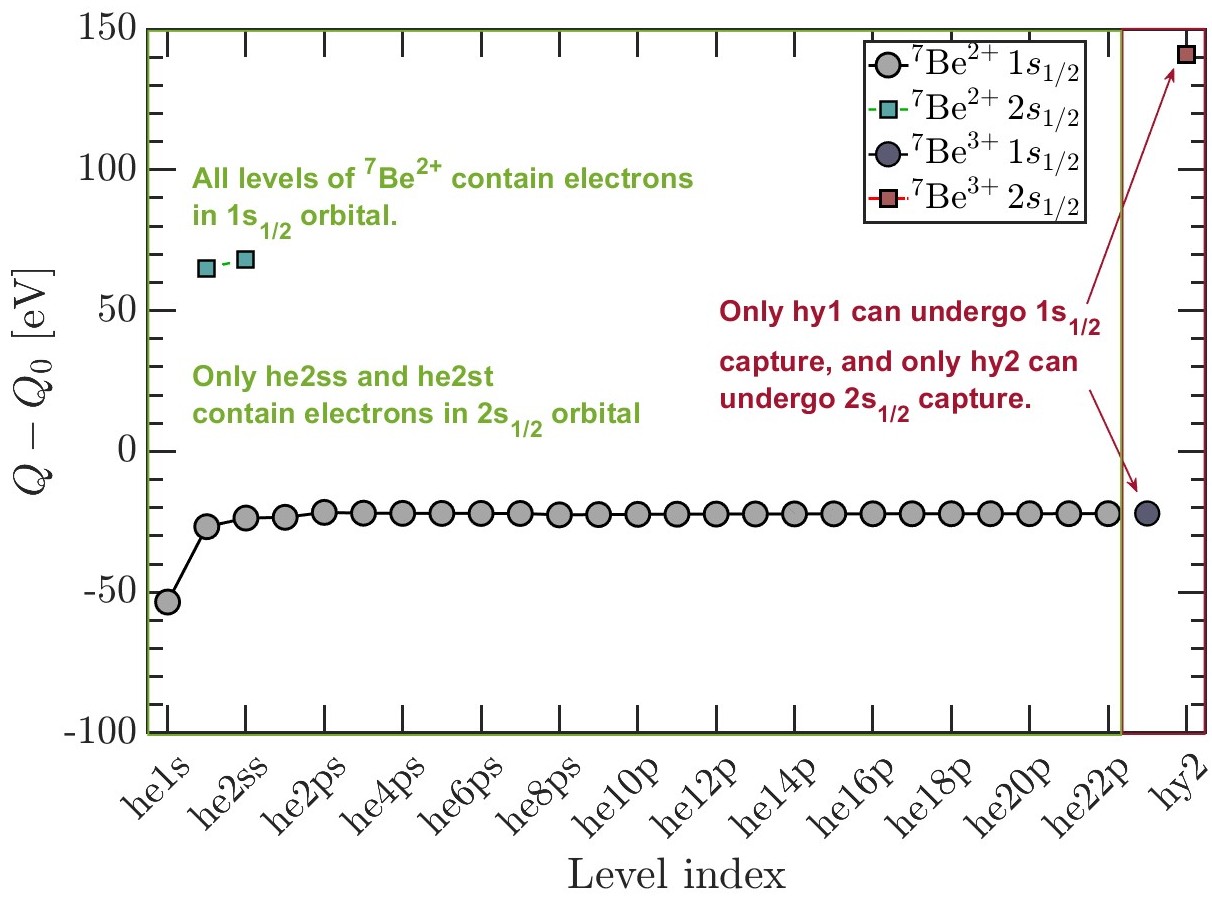}
      \caption{}
    \end{subfigure}
    \caption{$\Delta\epsilon^{ij,x}$ for different $j$ and $x=1s_{1/2},2s_{1/2}$ in (a) $^{7}\mathrm{Be}^{0,1+}$ and (b) $^{7}\mathrm{Be}^{2+,3+}$. No plot is shown for fully-ionised $^{7}\mathrm{Be}$ because it is void of electrons for capture. For sake of brevity, only a few levels in each ionisation stage are shown, but important levels are marked in the plots.}
    \label{Fig3}
\end{figure*}

\subsection{ERWF and Coulomb Amplitudes}
\label{fxgx}

The ERWFs $f_{x}$ and $g_{x}$ in Eq.~\ref{Eq2} describe the overlap between the nucleus and the atomic orbital \textit{x} from which the capture takes place. In the formalism of TY83, they are calculated as
\begin{align}
    \label{Eq11}
    f_{x}&=\frac{P_{\kappa}(r)}{r}\bigg|_{r=R} \\
    \label{Eq12}
    g_{x}&=\frac{Q_{\kappa}(r)}{r}\bigg|_{r=R}
\end{align}
where $R=R_{0}A^{1/3}$ is the phenomenological nuclear radius and $A$ is the mass number. The parameter $\kappa$ is related to the spin and angular momentum of the orbital $x$: $\kappa=l$ for $j=l-s$ and $\kappa=-(l+1)$ for $j=l+s$. The quantities $P_{\kappa}(r)$, $Q_{\kappa}(r)$ are the eigenfunctions of the radial component of the Dirac equations in a central Coulomb field generated by the nucleus and atomic electrons. 

As mentioned in Sec.\ref{Corr}, we use Coulomb amplitudes $\beta_{x}$ for calculating EC and BSBD rates in multi-electron ions. The $\beta_{x}$ are related to ERWFs and account for Coulomb screening by inner-shell electrons and the finite size of the nucleus. To calculate $\lambda^{*}(ij)$ in $^{7}$Be$^{0-2+}$, we use the tabulated $\beta_{x}$ from Ref.~\cite{Behrens1969} which contains pre-calculated $\beta_{1s_{1/2}}=9.444\times10^{-3}$ for $^{6}$Be, which we approximate for $^{7}$Be as well. Since the tables therein do not contain values for other orbitals, we use $\beta_{2s_{1/2}}=1.661\times10^{-3}$ from the BetaShape code~\cite{Mougeot2019,Mougeot2018}. These Coulomb amplitudes have been calculated for neutral atoms, but as electron screening decreases with ionisation, the use of the same $\beta_{x}$ in $^{7}$Be$^{1+}$ and $^{7}$Be$^{2+}$ leads to a slight underestimation in calculated capture rates. Future versions of our code will address this issue. 

For H-like $^{7}$Be$^{3+}$ instead, we directly use the analytical solutions to the Dirac equation from Burke and Grant~\cite{Burke1967} to calculate  $\beta_{1s_{1/2}}=10^{-2}$ and $\beta_{2s_{1/2}}=3.5\times10^{-3}$\footnote{The ERWFs for H-like ions are calculated on the nuclear surface $r=R$ (following the prescription of TY83), but the $\beta_{x}$ from Ref.~\cite{Behrens1969} and BetaShape~\cite{Mougeot2019,Mougeot2018} consider the overlap with the whole nuclear volume. The latter is physically correct, but the difference between the two is quite small, given the size of the Be nucleus. The situation may be different in heavier isotopes.}. These values are noticeably larger because the wavefunctions in hydrogenic ions overlap more strongly with the nucleus due to the absence of screening by other electrons. It is evident from the magnitude of $\beta_{x}$ that $K$-shell ($x=1s_{1/2}$) captures are the strongest, followed by $L$-captures. The $2s_{1/2}$ subshell dominates in the latter, and the contribution from the remaining orbitals can be neglected.    

\subsection{Shape Factor and Shell Occupancy}
\label{ShapeOcc}

The shape factor $S_{(m)x}$ is defined by the type of transition and orbital from which the electron is captured~\cite{TakahashiYokoi1983}. For allowed transitions such as in $^{7}$Be, $S_{(m)x}=1$ when capturing electrons from $x={1s_{1/2},2s_{1/2},2p_{1/2}}$ and $S_{m(x)}=0$ for $x=2p_{3/2}$. The shape factors are born from the selection rules and conservation of spin-parity between nucleon and lepton wavefunctions. 

The occupancy $\sigma_{x}$ is a number between $0$ and $1$ which describes the completeness of an orbital, with the former meaning completely empty and the latter meaning completely full. It is calculated according to the usual Pauli exclusion principles for atomic shell filling. To use Eq.~\ref{Eq5} identically for both EC and BSBD, $\sigma_{x}$ can be generalised as 
\begin{equation}
    \label{Eq13}
    (\sigma_{x})_{EC}=1-(\sigma_{x})_{BSBD}
\end{equation}
We calculate the occupancy of the $K$- and $L$-shell orbitals using the level descriptions of FLYCHK whose spectroscopic notations are often a direct representation of electron distribution in various orbitals (see Appendix~\ref{FLYCHK}).

\section{Results}
\label{Res}

\subsection{Configuration-Dependent Decay Rate}
\label{LevelDecay}

\begin{figure*}
\centering
    \begin{subfigure}{0.48\textwidth}
      \includegraphics[width=\textwidth]{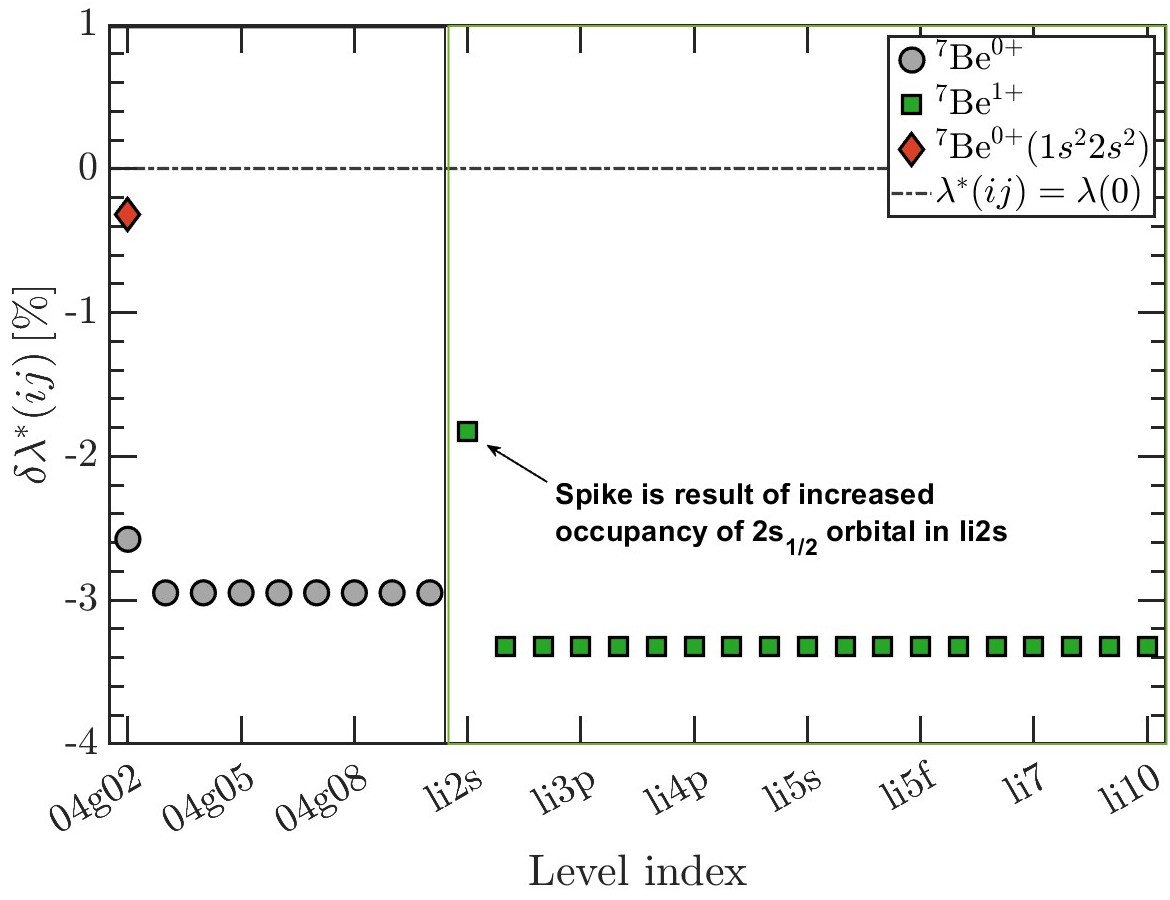}
      \caption{}
    \end{subfigure}
    \hfill
    \begin{subfigure}{0.49\textwidth}
      \includegraphics[width=\textwidth]{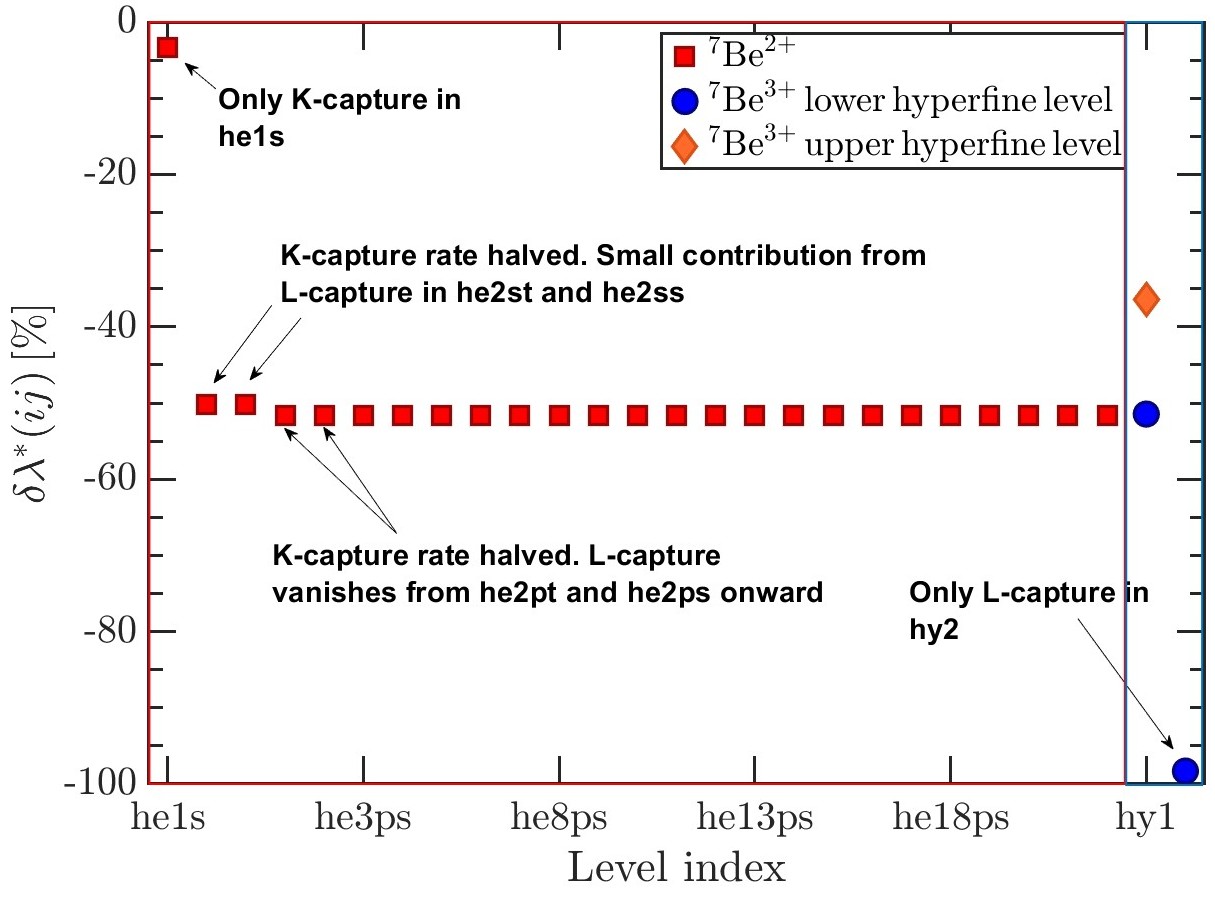}
      \caption{}
    \end{subfigure}
    \caption{Percentage change in decay rate with respect to neutral, ground state $^{7}$Be as a function of configuration $(ij)$ for (a) $i=0^{+},1^{+}$ and (b) $i=2^{+},3^{+}$. As in Fig.~\ref{Fig3}, only a few levels per ionisation stage are shown for brevity, but important levels are marked. The split at $\mathrm{hy}1$ is due to different contributions from $m=1$ to the total capture rate depending on whether the electron is in lower or upper hyperfine level. The red diamond in (a) refers to $\delta\lambda^{*}(ij)$ of the true ground state of $^{7}$Be - $1s^{2}2s^{2}$ configuration.}
    \label{Fig4}
\end{figure*}

By combining all the above quantities with Eqs.~\ref{Eq5} and \ref{Eq3}, we obtain the configuration-dependent EC rate $\lambda^{*}(ij)$ in $^{7}$Be. To demonstrate the effect of the ionic configuration better, we calculate the percentage change in EC decay rate $\delta\lambda^{*}(ij)$ as 
\begin{equation}
    \label{Eq14}
    \delta\lambda^{*}(ij)=\frac{\lambda^{*}(ij)-\lambda(0)}{\lambda(0)}\times100\%
\end{equation}
where $\lambda(0)=1.507\times10^{-7}\,\mathrm{s^{-1}}$ is the decay rate of neutral, ground state $^{7}$Be~\cite{Tilley2002}. The results are shown in Fig.~\ref{Fig4}.

It is evident that, in general, $^{7}$Be EC decay rate decreases with excitation and ionisation of the atom, with the drop being strongest in doubly and triply-ionised states. The main quantity driving this variation is the occupancy of the orbitals. The trends reported in Fig.~\ref{Fig4} can be explained as follows:
\begin{itemize}
    \item[$^{7}$Be$^{0,1+}$:] All bound levels contain a fully-occupied $K$-shell which dominates the capture process. Consequently, $\lambda^{*}(ij)$ do not differ much from $\lambda(0)$ ($|\delta\lambda^{*}(ij)|\leq3.5\%$). The small, additional suppression in $\lambda^{*}(ij)$ in excited levels is due to a $50\%$ and $100\%$ reduction in occupancy of the $L$-shell in $i=0^{+}$ and $1^{+}$, respectively.  
    \item[$^{7}$Be$^{2+}$:] A rapid drop from $\sigma_{1s_{1/2}}=100\%$ in $\mathrm{he}1s$ to $\sigma_{1s_{1/2}}=50\%$ in $\mathrm{he}2st$ results in an almost equal drop in decay rates. The rate decreases again going from $\mathrm{he}2ps$ to $\mathrm{he}3ps$ when $L$-captures switch from $2s_{1/2}$ to $2p_{1/2}$ orbitals, which are practically negligible. 
    \item[$^{7}$Be$^{3+}$:] $\lambda^{*}(ij)$ in $\mathrm{hy}1$ can take on two values depending on whether the electron is in the lower or upper hyperfine level. As described in Sec.~\ref{Corr}, in case of the former, the decay to the excited state of $^{7}$Li is blocked and therefore the total decay rate proceeds purely via the ground state transition, but in case of the latter, there is enhanced contribution from the $m=1$ transition. The final sharp drop by $\sim50\%$ in $^{7}$Be$^{3+}$ when going from $\mathrm{hy}1$ to $\mathrm{hy}2$ is explained by the excitation of the one attached electron from the $K$- to $L$-shell. 
\end{itemize}

It can be noted that predicted $\lambda^{*}(ij)$ is more suppressed in the ground state of $^{7}$Be$^{0+}$ (Fig.~\ref{Fig4}(a)) than expected ($\delta\lambda^{*}(ij)\sim2.6\%$). This can be explained by the imprecision in electronic configuration of neutral atoms in FLYCHK (see Appendix~\ref{FLYCHK} for more information). While the true ground state of $^{7}$Be$^{0+}$ is $1s^{2}2s^{2}$, the corresponding HULLAC configuration is $1s^{2}(2s2p)^{2}$ which reduces $\sigma_{2s_{1/2}}$ to $0.125$ and therefore suppresses $2s_{1/2}$-capture. The red diamond in Fig.~\ref{Fig4}(a) shows our model-prediction for the \textit{true} ground state of neutral beryllium and it can be observed that the rate is much closer to $\lambda(0)$. The small amount of mismatch may be attributed to the lack of atomic corrections reported in Subsec.~\ref{Corr}. The small spike in $\lambda^{*}(ij)$ at $\mathrm{li}2s$ compared to $04g02$ can also be attributed to the same imprecision - the electronic configuration of the former is $1s^{2}2s^{1}$ which results in $\sigma_{2s_{1/2}}=0.5$, whereas the same for the latter is $0.125$.

\begin{figure}
    \centering
    \includegraphics[width=0.96\linewidth]{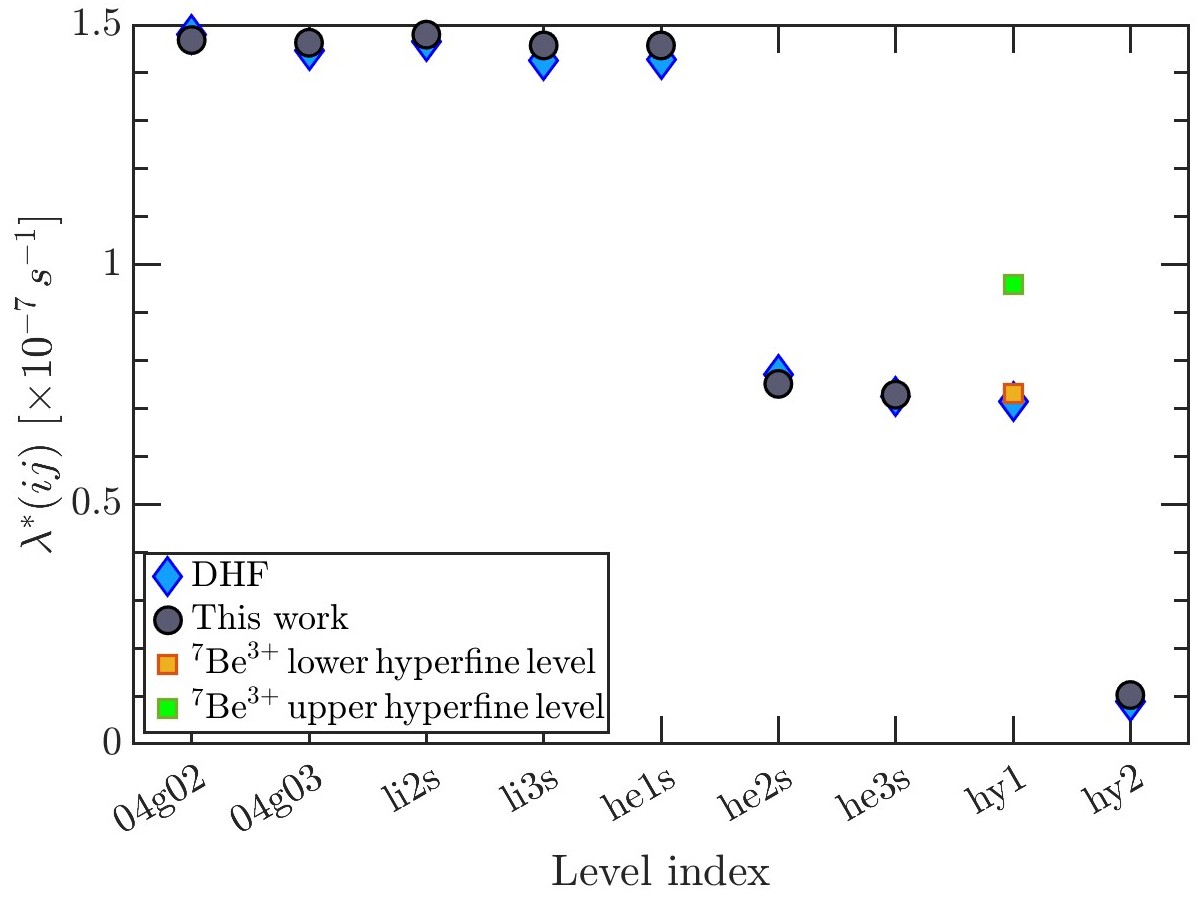}
    \caption{$^{7}$Be configuration-dependent $\lambda^{*}(ij)$ calculated using our model and DHF~\cite{Taioli2022,Simonucci2013,Morresi2018}.}
    \label{Fig5}
\end{figure}

The $\lambda^{*}(ij)$ calculated here are compared against independent results from the relativistic Dirac-Hartree-Fock (DHF) model~\cite{Taioli2022,Simonucci2013,Morresi2018} and the match between the two rates can also be appreciated in Fig.~\ref{Fig5}. Since the DHF model already includes exchange and overlap corrections, the positive match indicates that our model is fairly accurate in predicting the general trends of EC and BSBD in charged ions. The mismatch at $\mathrm{hy}1$ is due to the hyperfine splitting effect which has not been accounted for in the DHF model. When applied to neutral $^{7}$Be in \textit{true} ground state ($1s^{2}2s^{2}$), the model predicts $t_{1/2}=53.39\,\mathrm{d}$ and BR=$10.43\%$ which are close to the values reported in the ENSDF database~\cite{Tilley2002} ($t_{1/2}=53.22\pm0.06\,\mathrm{d}$ and BR=$10.44\%$). The $\lambda^{*}(ij)$ shown in Fig.~\ref{Fig5} will soon be validated against experimental data on EC in $^{7}$Be ions from the ERNA facility~\cite{Santonostaso2021}.

\subsubsection*{Validation in Other Nuclei}

The atomic component of the model is validated against EC and BSBD rates of certain radio-isotopes in select $(ij)$ configurations. We analysed the cases of $^{163}$Dy, $^{140}$Pr and $^{142}$Pm whose configuration-dependent decays have been experimentally studied in storage rings. The findings have been summarised in Tab.~\ref{Tab1}.

\begin{table}
\caption{\label{Tab1}
Comparison between $t_{1/2}$, $\lambda^{*}(ij)$, branching ratio and hyperfine enhancement of some isotopes as measured in storage rings or known from literature, and calculated by our model. Other independent predictions, if available, are also mentioned.}
\begin{ruledtabular}
\begin{tabular}{ccc}
\textbf{Isotope} &
\textbf{Measured} &
\textbf{Predicted} \\
\colrule
\\
$^{163}$Dy$^{66+}$  & $47^{+5}_{-4}\,\mathrm{d}$~\cite{Jung1992} & $49.77\,\mathrm{d}$  \\
(BSBD, $t_{1/2}$)  & & \textbf{Other}: $49.52\,\mathrm{d}$~\cite{Liu2021} \\\\
\hline
\\
$^{140}$Pr$^{0+}$  & $3.39(1)\,\mathrm{min}$~\cite{Nica2018} & $3.40\,\mathrm{min}$  \\
(EC+$\beta^{+}$, $t_{1/2}$)  & &  \\
$^{140}$Pr$^{57+}$  & $0.00147(7)\,\mathrm{s^{-1}}$~\cite{Litvinov2007} & $0.00153\,\mathrm{s^{-1}}$  \\
(EC, $\lambda^{*}(ij)$)  & &  \\
$^{140}$Pr$^{58+}$  & $0.00219(6)\,\mathrm{s^{-1}}$~\cite{Litvinov2007} & $0.00240\,\mathrm{s^{-1}}$  \\
(EC, $\lambda^{*}(ij)$)  & &  \\
$\frac{\lambda^{*}(\mathrm{H})}{\lambda^{*}(\mathrm{He})}$ & 1.49(8)\cite{Litvinov2007} & 1.57 \\\\
\hline
\\
$^{142}$Pm$^{0+}$  & $0.0039(5)\,\mathrm{s^{-1}}$~\cite{Winckler2009} & $0.0040\,\mathrm{s^{-1}}$  \\
(EC, $\lambda^{*}(ij)$)  & &  \\
$^{142}$Pm$^{59+}$  & $0.0036(1)\,\mathrm{s^{-1}}$~\cite{Winckler2009} & $0.0035\,\mathrm{s^{-1}}$  \\
(EC, $\lambda^{*}(ij)$)  & &  \\
$^{142}$Pm$^{60+}$  & $0.0051(1)\,\mathrm{s^{-1}}$~\cite{Winckler2009} & $0.0049\,\mathrm{s^{-1}}$  \\
(EC, $\lambda^{*}(ij)$)  & &  \\
$\frac{\lambda^{*}(\mathrm{H})}{\lambda^{*}(\mathrm{He})}$ & 1.44(6)\cite{Winckler2009} & 1.39 \\
\end{tabular}
\end{ruledtabular}
\end{table}

$^{140}$Pr: The isotope is unstable and undergoes decays through EC and $\beta^{+}$, with an accepted $t_{1/2}=3.39\pm0.01\,\mathrm{min}$ under neutral conditions~\cite{Nica2018}. The decay typically populates various levels of the daughter nucleus $^{140}$Ce, but we only consider the transitions $1^{+}\rightarrow0^{+}$, $1^{+}\rightarrow2^{+}$ and $1^{+}\rightarrow0^{+}$ populating, respectfully, the ground, first excited and second excited levels. The respective branching ratios are $99.39\%$, $0.27\%$ and $0.26\%$. Experiments in storage rings have measured $\lambda^{*}(ij)$ of $^{140}$Pr in fully-ionised, one-electron attached (hydrogen-like) and two-electron attached (helium-like) configurations~\cite{Litvinov2007}. We calculate the EC decay rate for helium-like $^{140}$Pr$^{57+}$ (summed over the aforementioned channels) and obtain $\lambda^{*}(ij)=1.53\times10^{-3}\,\mathrm{s^{-1}}$, in perfect agreement with measured rate $(1.47\pm0.07)\times10^{-3}\,\mathrm{s^{-1}}$. Our calculations use the Coulomb amplitude $\beta_{x}$ of $^{136}$Pr from Ref.~\cite{Behrens1969}. We also calculate the corresponding rate in $^{140}$Pr$^{58+}$, factoring in the hyperfine splitting for each of the transitions. Using the formulation in Ref.~\cite{Patyk2008} and assuming that the electron resides in the lower hyperfine level of $F_{i1}=1-1/2$, the respective coefficients come out to be $F(m)=3$, $0$ and $3$. We therefore obtain $\lambda^{*}(ij)=2.4\times10^{-3}\,\mathrm{s^{-1}}$ for hydrogen-like $^{140}$Pr$^{58+}$, slightly higher than the measured value of $(2.19\pm0.06)\times10^{-3}\,\mathrm{s^{-1}}$. However, the ratio of EC decay rate in $^{140}$Pr$^{58+}$ and $^{140}$Pr$^{57+}$ is calculated to be $\lambda^{\ast}(\mathrm{H})/\lambda^{\ast}(\mathrm{He})=1.57$ which is within the error limits of experimentally measured $1.49\pm0.08$. The reason for the overestimation is that our model does not yet take into account electron correlations in helium-like ions which have a small effect on $F(m)$. The authors in Ref.~\cite{Litvinov2007} also note that $\beta^{+}$ decay rates are fairly invariant over different atomic configurations, and on adding their measured $\lambda^{*}_{\beta^{+}}$ to our calculated $\lambda^{*}_{EC}$ for neutral $^{140}$Pr, we obtain $t_{1/2}=3.40\,\mathrm{min}$ which also in full agreement with the aforementioned half-life of $^{140}$Pr$^{0+}$.

$^{142}$Pm: The isotope is similar in nature to $^{140}$Pr. It is also unstable, and undergoes EC + $\beta^{+}$ decays to $^{142}$Nd with $t_{1/2}=40.5\pm0.5\,\mathrm{s}$~\cite{Johnson2011}. The ground state of the parent at $1^{+}$ is coupled to various daughter levels, the strongest being the transition to the ground state $0^{+}$ ($\mathrm{BR}=96.44\%$). We use Eq.~\ref{Eq5} to calculate EC decay rates in neutral, helium-like and hydrogen-like $^{142}$Pm ions and compare the results with measured values from storage rings~\cite{Winckler2009}. We use the pre-calculated $\beta_{x}$ of $^{141}$Pm from Ref.~\cite{Behrens1969} and implement hyperfine splitting, once again assuming that the electron only populates the lower hyperfine level $F_{i1}=I_{i}-1/2$. We obtain $\lambda^{\ast}(ij)=4.0\times10^{-3}\,\mathrm{s^{-1}}$, $3.5\times10^{-3}\,\mathrm{s^{-1}}$ and $4.9\times10^{-3}\,\mathrm{s^{-1}}$ for $^{142}$Pm$^{0+}$, $^{142}$Pm$^{59+}$ and $^{142}$Pm$^{60+}$, respectively. The corresponding measured values are $(3.9\pm0.5)\times10^{-3}\,\mathrm{s^{-1}}$, $(3.6\pm0.1)\times10^{-3}\,\mathrm{s^{-1}}$ and $(5.1\pm0.1)\times10^{-3}\,\mathrm{s^{-1}}$, and therefore there is a perfect match between our model-prediction and experimental data. In addition, our calculated ratio $\lambda^{\ast}(\mathrm{H})/\lambda^{\ast}(\mathrm{He})=1.39$ which is well within the uncertainty of the measured value of $1.44\pm0.06$.  

$^{163}$Dy: The isotope is stable under neutral conditions, but when fully-ionised, it undergoes BSBD with a measured $t_{1/2}=47^{+5}_{-4}\,\mathrm{d}$~\cite{Jung1992}. On applying Eq.~\ref{Eq5} to $^{166}$Dy$^{66+}$ and using a single transition marked by $\mathrm{log}ft=4.99$, we calculate $t_{1/2}=49.77\,\mathrm{d}$, which is not only within the experimental error limits, but also the same as calculated in Chang and coworkers~\cite{Liu2021} ($t_{1/2}=49.52\,\mathrm{d}$).

\subsection{In-Plasma Decay Rate}
\label{PlasmaDecay}

The configuration-dependent decay rates $\lambda^{*}(ij)$ can be converted into the in-plasma decay rates through the equation
\begin{equation}
    \label{Eq15}
    \lambda^{*}=\sum_{ij}p_{ij}\lambda^{*}(ij)
\end{equation}
where $p_{ij}$ indicates the probability of finding the ion in the $(ij)$-configuration. The purpose of the plasma component of our model is to customise the formalism to evaluate $p_{ij}$ according to the geometry and nature of the plasma.

For \textit{uniform} plasmas (homogeneous and isotropic), we use FLYCHK to obtain $p_{ij}$ in a large range of $n_{e}$ and $kT_{e}$ values. The calculations can be run in both LTE and NLTE modes, which allows us to explore plasma ion dynamics under a variety of situations and study their effect on $\lambda^{*}$.

\begin{figure}
    \centering
    \includegraphics[width=0.97\linewidth]{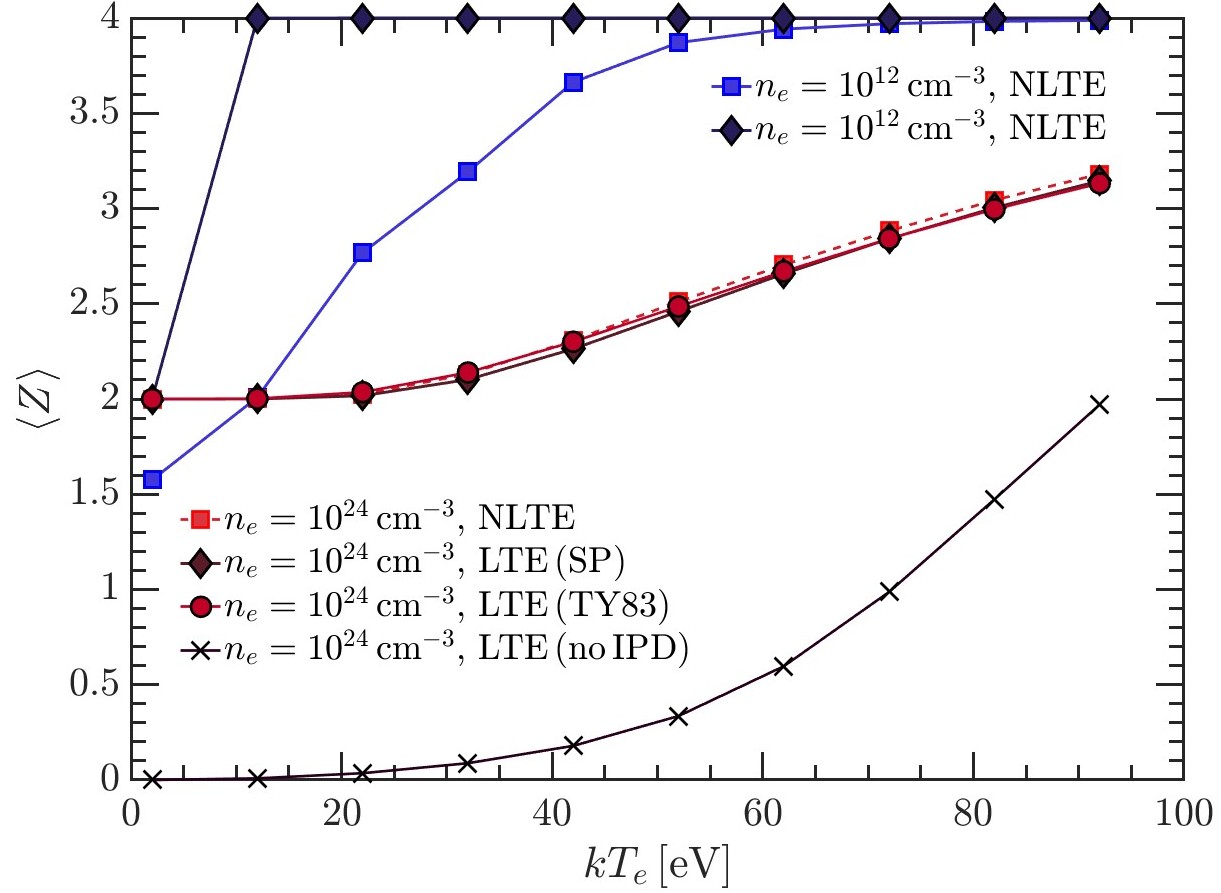}
    \caption{$\langle Z\rangle$ of $^{7}$Be as a function of $kT_{e}$ for $n_{e}=10^{12}\,\mathrm{cm^{-3}}$ (blue markers) and $n_{e}=10^{24}\,\mathrm{cm^{-3}}$ (red markers). The crossed, continuous black line represents a high-density plasma with no IPD correction.}
    \label{Fig6}
\end{figure}

Fig.~\ref{Fig6} shows the mean charge $\langle Z\rangle$ of $^{7}$Be ions as a function of plasma electron temperature between $2\,\mathrm{eV}$ and $92\,\mathrm{eV}$, under conditions of low and high density. The plasma is taken to be purely composed of $^{7}$Be and electrons. The dark blue diamonds and the blue squares represent $\langle Z\rangle$ calculated by FLYCHK under LTE and NLTE conditions, respectively. It can be observed that the trends converge only at high $kT_{e}$ - at low temperatures, LTE predictions overestimate the presence of $^{7}$Be$^{4+}$. LTE can be considered as a special case of NLTE under conditions of high collisionality. The plot serves to prove that at low $n_{e}$ such as that which will be found in PANDORA, TY83 cannot be directly applied because in its original form, it is only applicable to LTE plasmas.

The set of red curves in Fig.~\ref{Fig6} depicts the mean charge of $^{7}$Be at $n_{e}=10^{24}\,\mathrm{cm^{-3}}$. The red squares represent $\langle Z\rangle$ under NLTE conditions, whereas the dark red diamonds showcase the same under LTE. Both sets of data are obtained from FLYCHK. It can be immediately noted that contrary to the low density case, NLTE and LTE $\langle Z\rangle$ are identical at high density. This is because the plasma is already sufficiently collisional, even at low temperatures.

In dense plasmas, the electron cloud screens the positive Coulomb field of the nucleus and suppresses the energy of bound levels. This so-called ionisation potential depression (IPD) strongly modifies the energies of atomic levels as a function of $n_{e}$ and $kT_{e}$, and can auto-ionise the atoms by depressing out low charge states. To emphasise the impact of IPD at high $n_{e}$, Fig.~\ref{Fig6} also shows a comparison between three different sets of LTE mean charge. The FLYCHK calculations (dark red diamonds) implement the Stewart-Pyatt (SP) model of IPD~\cite{SP1965} 
\begin{equation}
    \label{Eq16}
    \Delta E_{i}=2.16\times10^{-7}\frac{(i+1)}{r_{s}}\bigg(\bigg(1+\bigg(\frac{r_{d}}{r_{s}}\bigg)^{3}\bigg)^{2/3}-\bigg(\frac{r_{d}}{r_{s}}\bigg)^{2}\bigg)
\end{equation}
where $\Delta E_{i}$ is the reduction in energy of the $i^{\mathrm{th}}$ charge state (in $\mathrm{eV}$) and $r_{d}$, $r_{s}$ are respectively the Debye sphere and ion sphere radii. These latter depend on the plasma parameters $n_{e}$, $kT_{e}$ and the mean charge of the ions. The corresponding data points are labelled as \textbf{LTE (SP)} in the legend. The red circles showcase $\langle Z\rangle$ calculated using the Saha equation and IPD taken from the approximate interpolation formula in TY83 (labelled \textbf{LTE (TY83)})
\begin{equation}
    \label{Eq17}
    \begin{split}
    \log(\Delta E_{i})=[d_{1}\log(T_{7})+d_{2}]\log(i+1)+ \\
    [d_{3}(\log(T_{7})^{2}+d_{4}\log(T_{7})+d_{5}]    
    \end{split}
\end{equation}
where $T_{7}$ is electron temperature in $10^{7}\,\mathrm{K}$, $d_{1-5}$ are coefficients involving $n_{e}$, and $\Delta E_{i}$ is in $\mathrm{keV}$ (see Ref.~\cite{TakahashiYokoi1983} for details). The crossed black line shows the same Saha implementation, this time with \textit{no} IPD. It can be easily noted that the no-IPD mean charge is significantly lower than the IPD-prediction at all $kT_{e}$. This is because at high $n_{e}$, recombination dominates over ionisation and $^{7}$Be does not make it beyond the first two charge states. However, lowering of the ionisation potential causes these first few charge states to get depressed out ($\Delta E_{i}>E_{i}$ for $i=0^{+},1^{+}$), resulting in a charge state distribution (CSD) which only includes $i=2^{+},3^{+}$ and $4^{+}$ and therefore a much higher $\langle Z\rangle$. This trend is true for both SP and TY83 formulations, with only negligible differences at high $kT_{e}$. The CSD of $^{7}$Be predicted by the two formulations deviates significantly at $n_{e}>10^{25}\,\mathrm{cm^{-3}}$ and low $kT_{e}$.

\subsubsection{Solar Plasma}
\label{SolarDecayRate}

To verify the plasma component of the model and portray its validity in a wider parameter space, the EC decay rate of $^{7}$Be is calculated in the solar interior. We use Eq.~\ref{Eq15}, assuming that $\lambda^{*}(ij)$ remains invariant inside the high-$n_{e}$ stellar plasma. We first calculate the CSD $p_{i}$ of $^{7}$Be in the solar interior using the Saha equation with SP-IPD, as used in FLYCHK. 

\begin{figure*}
    \includegraphics[width=0.8\textwidth]{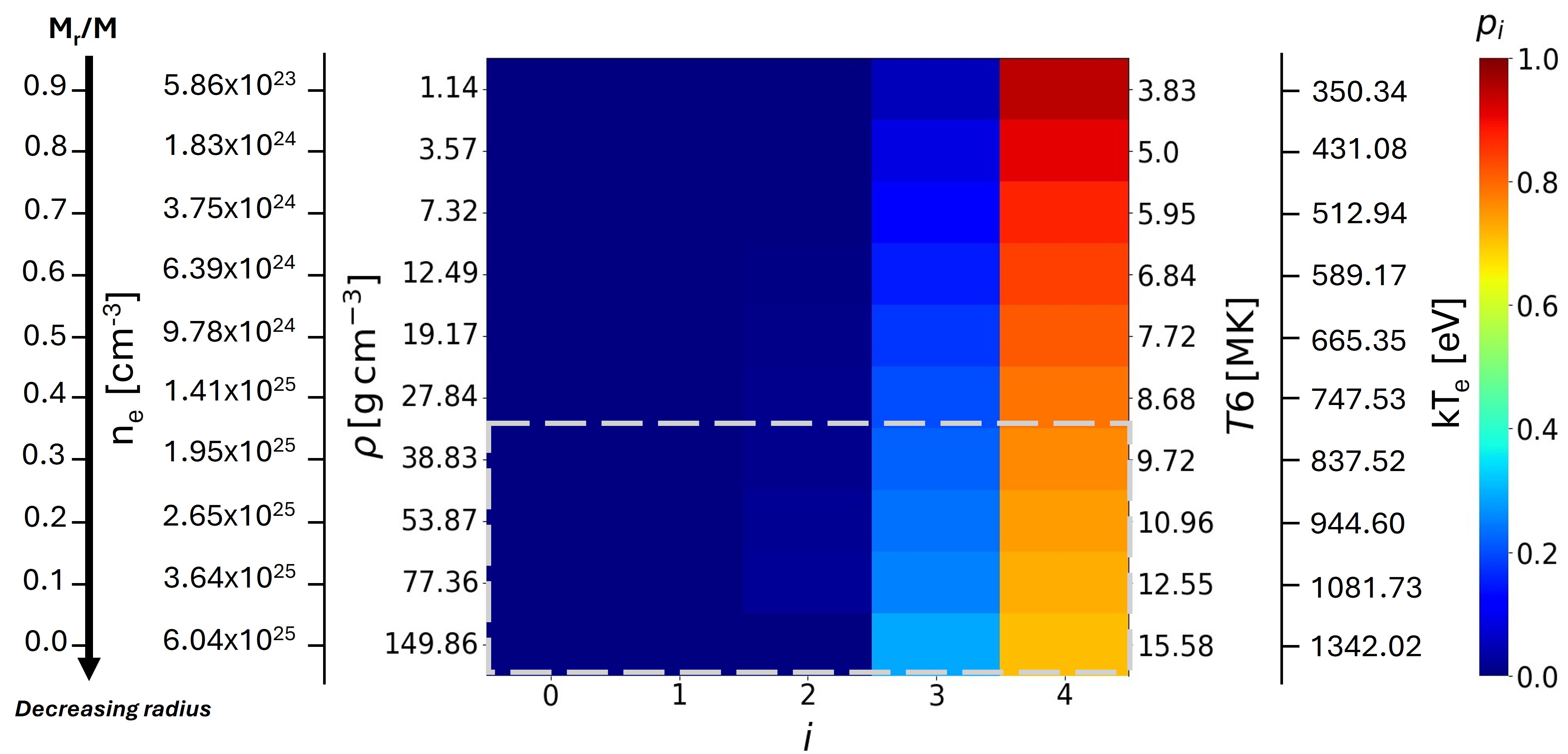}
    \caption{$^{7}$Be $p_{i}$ as a function of solar mass coordinate $M_{r}/M$.}
    \label{Fig7}
\end{figure*}

Fig.~\ref{Fig7} shows the resultant $p_{i}$ as a function of the mass coordinate in the solar core. The mass coordinate is defined as the ratio $M_{r}/M$ where $M$ is the mass of the sun and $M_{r}$ is the mass enclosed within a sphere of radius $r$. It decreases from $M_{r}/M=0.9$ at the top of the vertical axis to $M_{r}/M=0$ at the bottom, denoting a descent along the radius into the core (the smaller the mass coordinate, the closer one is to the centre of the sun). The horizontal axis shows the charge states of $^{7}$Be whereas the (inner) left and right vertical axes show, respectively, the mass density $\rho$ (in $\mathrm{g\,cm^{-3}}$) and temperature of the plasma $T_{6}$ (in $10^{6}\,\mathrm{K}$) at that mass coordinate. To facilitate analysis, a secondary (outer) vertical axis is added to each side of the plot, denoting the plasma electron density $n_{e}$ corresponding to $\rho$, and electron temperature $kT_{e}$ corresponding to $T_{6}$. The quantities are related to each other as 
\begin{align}
    \label{Eq18}
    n_{e}\,[\mathrm{cm^{-3}}]&=\frac{\rho}{m_{p}\mu_{e}} \\
    \label{Eq19}
    kT_{e}\,[\mathrm{eV}]&=\frac{T_{6}\times10^{6}}{11606}
\end{align}
where $m_{p}$ is the rest mass of the proton (in $\mathrm{g}$) and $\mu_{e}$ is the mean molecular weight of the electron. The latter is calculated as $\mu_{e}\approx2/(1+X_{H})$ where $X_{H}$ is the mass fraction of hydrogen. $\rho$, $T_{6}$ and $X_{H}$ for different mass coordinates were taken from the solar composition deduced from photospheric observations by Magg \textit{et al}~\cite{Magg2022}, compiled in the updated standard solar model (SSM) database by Serenelli and Herrera~\cite{SSM2023}.  

The dashed, grey box in Fig.~\ref{Fig7} encloses the parameter space of the inner $30\%$ of the solar interior, where $30\leq\rho\leq 150\,\mathrm{g\,cm^{-3}}$ and $9\leq T_{6}\leq16\,\mathrm{MK}$. As can be noted, at $M_{r}/M=0.1$ ($\rho=77.36\,\mathrm{g\,cm^{-3}}, T_{6}=12.55\,\mathrm{MK}$), $p_{i=3^{+}}\sim29\%$. This indicates a non-zero contribution of orbital EC from $^{7}$Be$^{3+}$, which has been noted in literature~\cite{Iben1967,Bahcall1969,Gruzinov1997,Taioli2022}.

\begin{figure}
    \centering
    \includegraphics[width=0.99\linewidth]{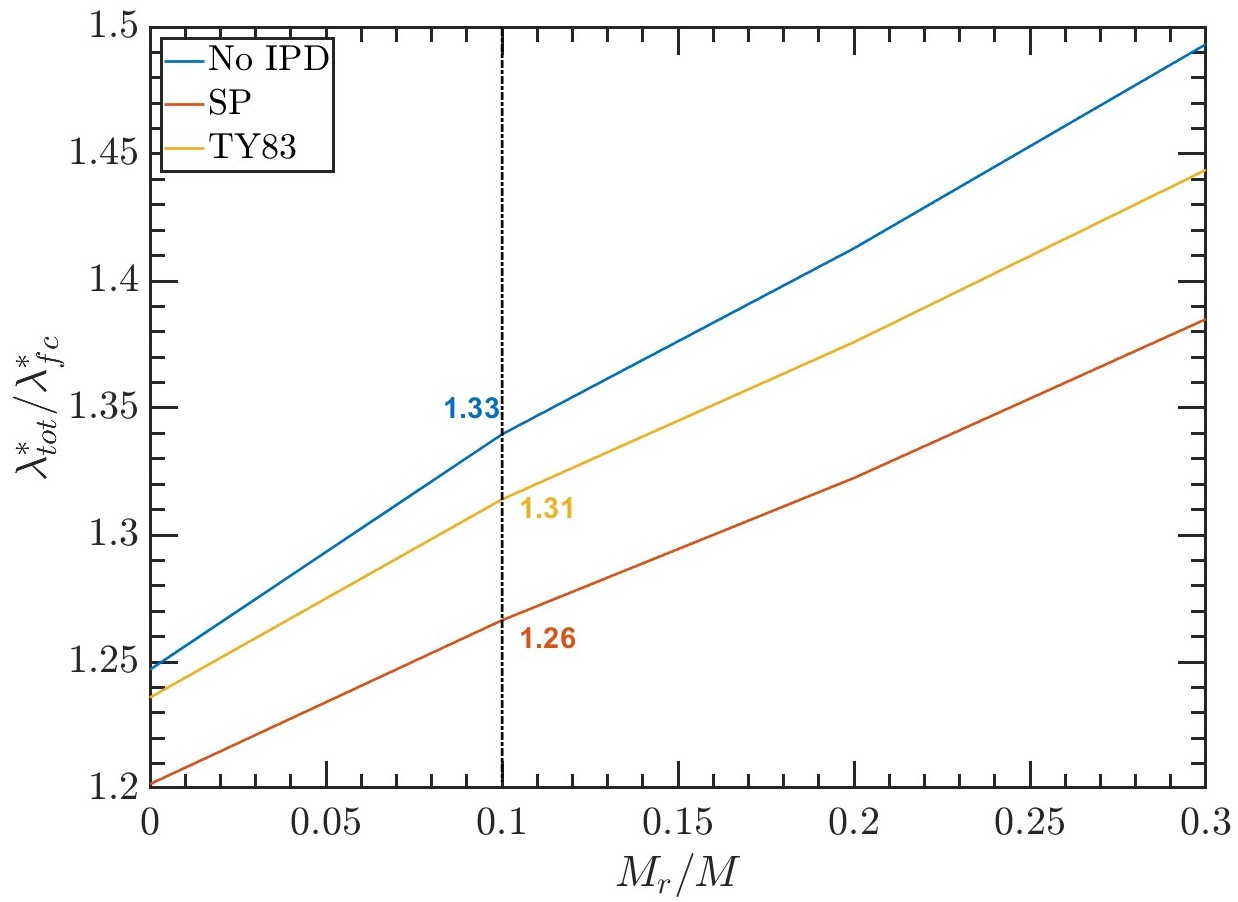}
    \caption{Enhancement in capture rate of $^{7}$Be at $M_{r}/M\leq 0.3$ as predicted by Saha equation with no-IPD, TY83-IPD~\cite{TakahashiYokoi1983} and SP-IPD~\cite{SP1965}.}
    \label{Fig8}
\end{figure}

To quantify the impact of non-zero orbital EC from $i=3^{+}$, the total decay rate of $^{7}$Be in the solar interior can be calculated as
\begin{equation}
    \label{Eq20}
    \lambda^{*}_{tot}=\sum_{i=0^{+}}^{3^{+}}p_{i}\lambda^{*}(i)+\lambda^{*}_{fc}
\end{equation}
where $\lambda^{*}(i)$ is the configuration-dependent EC rate of $^{7}$Be in $i^{+}$ charge state and ground excitation level, and $\lambda^{*}_{fc}$ is the free electron capture rate. The latter was directly taken from the approximate formulation of Bahcall~\cite{Bahcall1969}, valid in the range $10\leq T_{6}\leq16\,\mathrm{MK}$:
\begin{equation}
    \label{Eq21}
    \lambda^{*}_{fc}=4.62\times10^{-9}(\rho/\mu_{e})T_{6}^{-1/2}[1+0.004(T_{6}-16)]
\end{equation}
Eq.~\ref{Eq20} is used to calculate the ratio $\lambda^{*}_{tot}/\lambda^{*}_{fc}$ in the region $M_{r}/M\leq0.3$. 
Fig.~\ref{Fig8} shows the ratio as a function of the mass coordinate in the solar interior, as predicted by three different LTE models - pure Saha equation with no IPD, Saha equation with IPD from TY83, and IPD from SP (FLYCHK). The vertical line indicates the mass coordinate $M_{r}/M=0.1$ at which most of the solar neutrinos are generated. It can be noted that both no-IPD and TY83 overestimate the enhancement of $^{7}$Be decay, but the model proposed in this work predicts $\lambda^{*}_{tot}/\lambda^{*}_{fc}\sim1.26$. This value is in accordance with the calculation of Iben and Bahcall who predicted a similar enhancement ($\lambda^{*}_{tot}/\lambda^{*}_{fc}\sim1.25$) using a Debye-H\"{u}ckel screening model~\cite{Iben1967,Bahcall1969,Gruzinov1997}.

\subsubsection{Laboratory Magnetoplasma}
\label{PANDORADecayRate}

The model is then applied to the expected parameter space of the PANDORA laboratory magnetoplasma trap to predict $\lambda^{*}$ as a function of $n_{e}$ and $kT_{e}$. 

\begin{figure*}
\centering
    \begin{subfigure}{0.49\textwidth}
      \includegraphics[width=\textwidth]{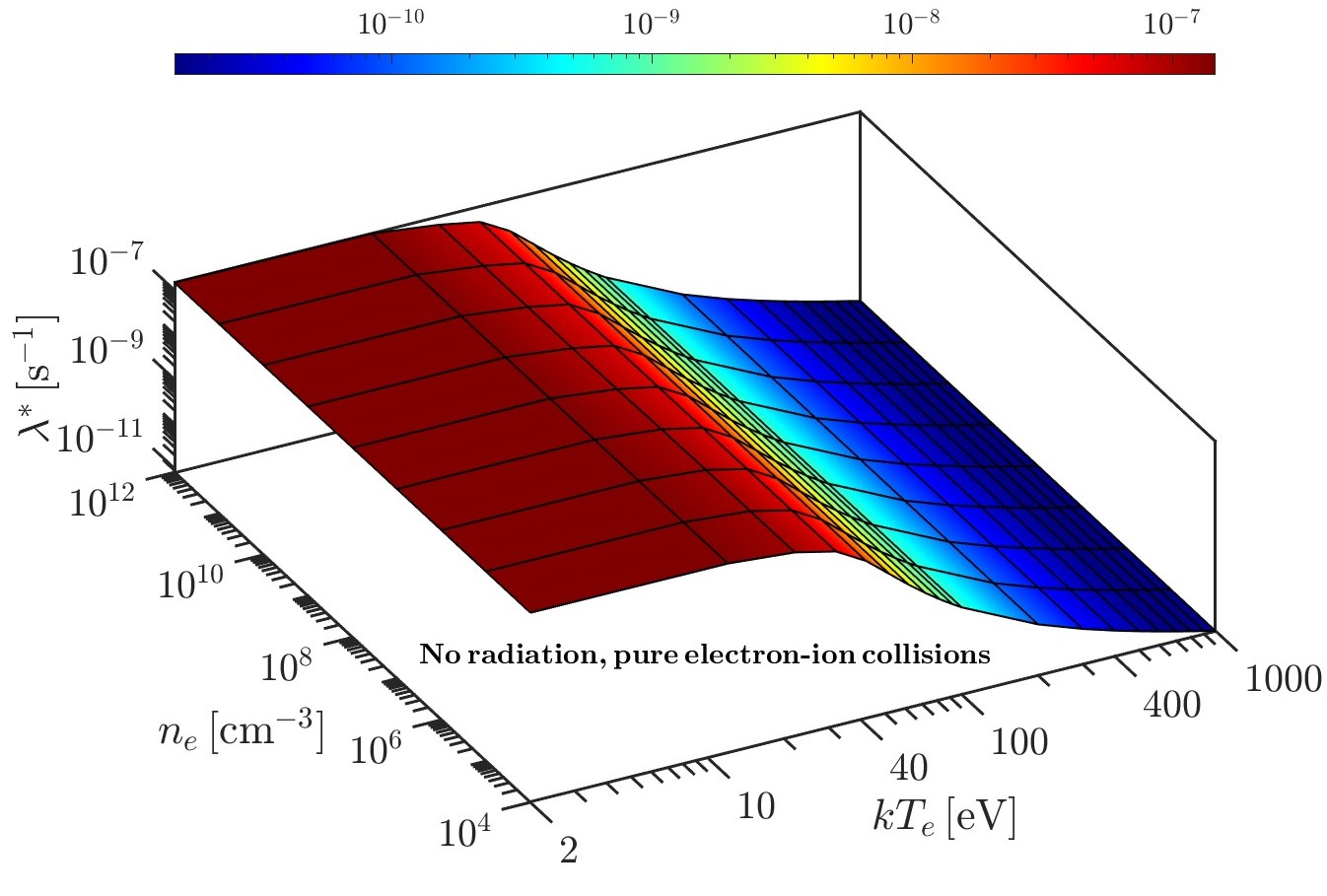}
      \caption{}
    \end{subfigure}
    \hfill
    \begin{subfigure}{0.49\textwidth}
      \includegraphics[width=\textwidth]{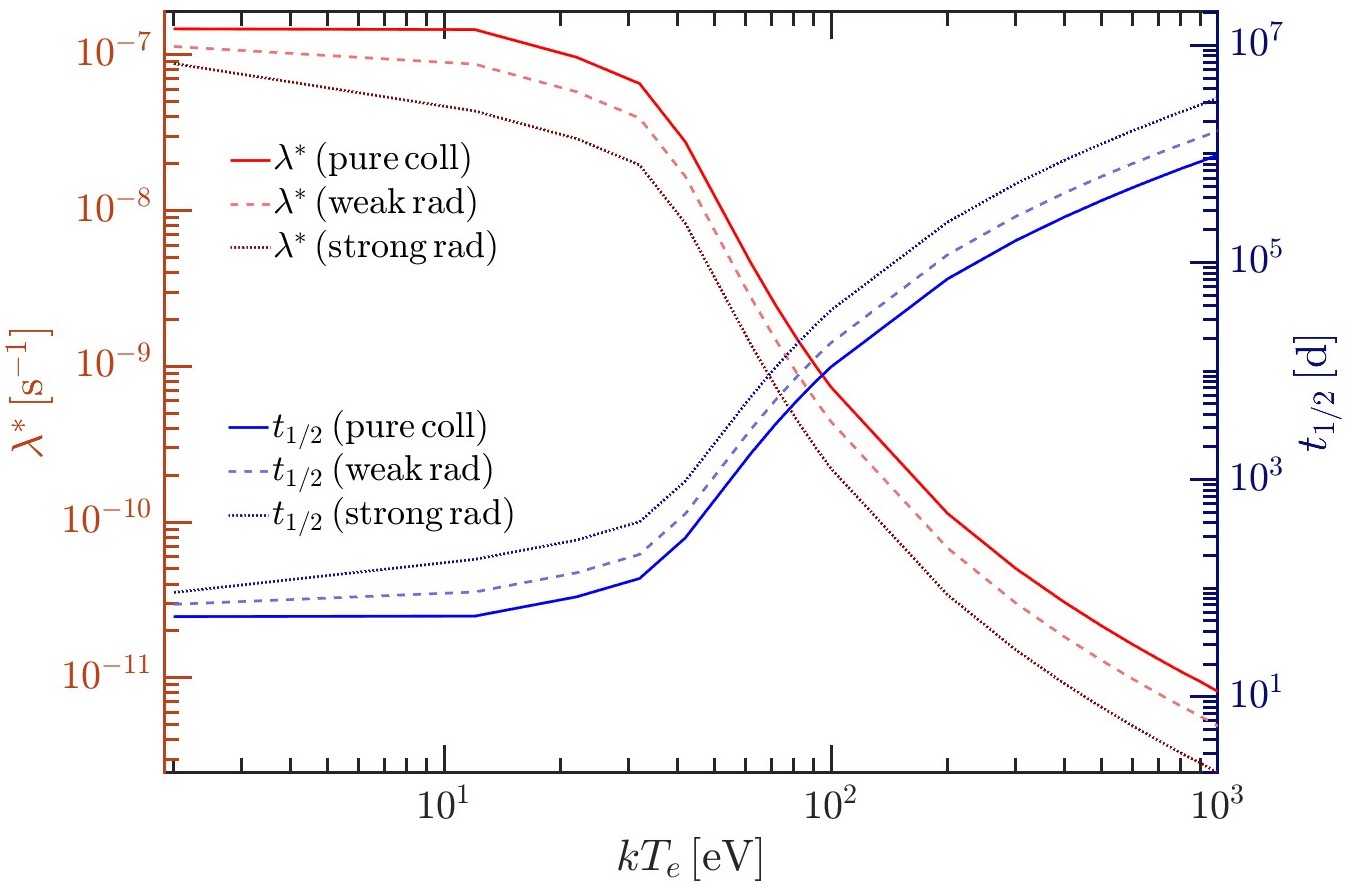}
      \caption{}
    \end{subfigure}
    \caption{$\lambda^{*}$ as a function of $n_{e}$ and $kT_{e}$ in the realistic parameter space of PANDORA and (b) variation of $\lambda^{*}$ and $t_{1/2}$ of $^{7}$Be as a function of $kT_{e}$ for $n_{e}=10^{12}\,\mathrm{cm^{-3}}$ under different radiation field strengths.}
    \label{Fig9}    
\end{figure*}

We run FLYCHK simulations for $n_{e}\in[10^{4},10^{12}]\,\mathrm{cm^{-3}}$ and $kT_{e}\in[1,1000]\,\mathrm{eV}$ under NLTE conditions. To simulate a more realistic scenario, the plasma is taken to be primarily composed of electrons and buffer ions of helium, with $^{7}$Be being a minority species at a concentration $1\%$ in number density. The resultant $p_{ij}$ are used to calculate $\lambda^{*}$ according to Eq.~\ref{Eq15}.  Fig.~\ref{Fig9}(a) shows the variation of $\lambda^{*}$ as a function of $n_{e}$ and $kT_{e}$. It can be easily noted that at low $n_{e}$, $\lambda^{*}$ is density-invariant, and depends only on the temperature $kT_{e}$. Unsurprisingly, the decay rate drops with increasing temperature due to the production of $^{7}$Be$^{3+}$ and $^{7}$Be$^{4+}$ ions which strongly suppress orbital EC. At $kT_{e}\sim100\,\mathrm{eV}$, $\lambda^{*}$ can already be expected to drop by two orders of magnitude with respect to the neutral rate. Such a temperature will be easily achievable under normal operating conditions in PANDORA. Further increase in $kT_{e}$ will only increase the amount of $i=4^{+}$ ions, thereby decreasing $\lambda^{*}$ even more. It should be mentioned that free electron capture is negligible at such low $n_{e}$ and hence $\lambda^{*}_{tot}$ will asymptotically approach $0$ as temperatures keep rising.

An important fact to underline here is the role of radiation. The CSD and atomic excitation of ions in NLTE plasmas are calculated using a collision-radiative (CR) model. A CR model typically takes the form of a 2D matrix whose elements represent the transition rate between different excitation levels and charge states of the ion. The transition rate is a sum of three principal reactions: electron-ion collisions, ion-ion collisions, and radiation-induced processes. The $p_{ij}$ factors calculated by FLYCHK and used in Fig.~\ref{Fig9}(a) are obtained from a \textit{reduced} CR model which only includes electron-ion collisions. The algorithm implemented by FLYCHK does not account for inter-ion collisions, and while radiative processes such as photoexcitation and photoionisation \textit{are} present in the model, their implementation requires defining the plasma opacity (particularly in the visible and UV range). Work is currently underway to calculate and measure the opacity of low-density magnetoplasmas \cite{Pidatella2021,Pidatella2022}. In the meantime however, the impact of the radiation field on $\lambda^{*}$ can be explored through its action in equilibrating the population of ground and excited levels within a charge state, specifically in He- and H-like ions.

Fig.~\ref{Fig9}(b) shows the variation of $\lambda^{*}$ (left vertical axis) and its corresponding $t_{1/2}$ (right vertical axis) as a function of $kT_{e}$ for $n_{e}=10^{12}\,\mathrm{cm^{-3}}$. The continuous lines represent ions subject to only electron-collisions, whereas the light and dark dashed lines refer to, respectively, weak and strong radiation fields. The more opaque the plasma, the more radiation is trapped and larger the photoexcitation rates. The added contribution of radiative transitions can consequently increase the population of excited levels with respect to the ground. The "weak radiation" regime here is a hypothetical case which is assumed to equilibrate bound excited levels in $^{7}$Be$^{2+}$ and $^{7}$Be$^{3+}$ in a ratio $40\%:60\%$ with respect to the ground, whereas the "strong radiation" equilibrates them in the ratio $70\%:30\%$. These assumptions are arbitrary and merely illustrative in nature, but their purpose is to show that depending on the intensity of the self-generated radiation field in the magnetoplasma (particularly of photons in the energy range $100-200\,\mathrm{eV}$), a lower $\lambda^{*}$ may be measured under otherwise identical conditions due to added suppression of EC-rates from bound excited levels. This fact is evident in Fig.~\ref{Fig9}(b) where the curves of $\lambda^{*}$ under radiative effects fall below that of pure-collisional plasmas. Radiative processes can therefore play a strong role in low-density, NLTE plasmas and the model described in this work can be suitably adapted to take them into account.

\section{Conclusion}
\label{Conc}

In summary, we have developed a general model of in-plasma $\beta$ decay which can be identically applied to a wide plasma parameter space, from dense stars to rarified laboratory magnetoplasmas. The strength of the model lies in its relatively simple formulation which allows predicting EC and BSBD rates in isotopes of varying masses in all charge states and excitation levels with a fair amount of accuracy. The separation of atomic and plasma components maintains versatility when dealing with different environments, and generates more observables which can be tested with different facilities. The model has been successfully tested with measurements from highly-charged ions of $^{163}$Dy, $^{140}$Pr and $^{142}$Pm as well as with the solar EC decay rate of $^{7}$Be. The application of the model to PANDORA suggests that $^{7}$Be EC decay rates could decrease by several orders of magnitude even at nominal operating conditions of the trap, and that the opacity of the plasma can strongly impact the rate by equilibrating excited states of H- and He-like ions.

The plots of Fig.~\ref{Fig9} provide a first estimate of $\lambda^{*}$ as a function of $n_{e}$ and $kT_{e}$, but to fulfill the objective of fully benchmarking TY83 in PANDORA, one needs to account for specific properties of ECR plasmas. Therefore, we have coupled our model with a Particle-in-Cell Monte Carlo (PIC-MC) code \cite{Mishra2021} to simulate $p_{ij}$ in the real PANDORA trap. We have already obtained first maps of 3D \cite{Naselli2025,Pidatella2024}, space-resolved in-plasma decay rates awaiting experimental verification by PANDORA. We are also working on calculating plasma opacity in the visible-IR-UV range which will help improve the model and assess the impact of radiation on the decay rates. These results will be published soon.

We are working on other upgrades to improve the applicability of our model. First and foremost is inclusion of the various corrections listed in Subsec.~\ref{Corr} and a more rigorous evaluation of Coulomb amplitudes. This would allow extending the Eq.~\ref{Eq5} to all charge states. We also intend to investigate in-plasma \textit{continuum} captures and $\beta^{+/-}$ decays, complementary to the work by Gupta and coworkers~\cite{Gupta2019,Gupta2023}. This extension involves including excited nuclear states in the parent and daughter atoms and assessing their impact on the modification of $t_{1/2}$. Since these additional nuclear couplings lack literature data on their comparative half-lives, we are looking into shell model and \textit{ab-initio} calculations to obtain better precision on $\mathrm{log}ft$ values, in line with the suggestions by Chang and coworkers~\cite{Liu2021}.

\begin{acknowledgments}
The authors wish to thank INFN for the support through the project PANDORA\_Gr3 funded by $3^{rd}$ Nat. Sci. Comm. A.P. would like to acknowledge the financial support from the MUR-PNRR Project PE0000023-NQSTI, financed by the European Union (NextGeneration EU). S.T. would like to acknowledge the European Union under grant agreement no. 101046651 (MIMOSA) for this action. The authors also warmly acknowledge the useful discussions with Chang Xu, Shuo Liu and Kohji Takahashi on the topic of ERWFs, and with Yury Litvinov on measurment of hyperfine interactions in hydrogenic ions.
\end{acknowledgments}


\bibliography{apssamp}

\appendix

\section{Hyperfine Splitting}
\label{HyperSplit}

In case of hydrogen-like $^{7}$Be$^{3+}$ in ground state (electronic configuration $1s^{1}$), the interaction between the nucleus and the $K$-shell splits the latter into two hyperfine levels with total spin $F_{i1}=3/2-1/2$ and $F_{i2}=3/2+1/2$. The energy difference between the two levels is denoted by $\Delta$ and is a small but unknown number. Owing to the negative nuclear magnetic moment of $^{7}$Be (about $-1.3\mu_{N}$~\cite{Rose1937,Navratil1998,Doma2007}), $F_{i1}$ is larger in energy and is termed the upper hyperfine level. The daughter system is a fully-ionised $^{7}$Li$^{3+}$ ion coupled to a single, free neutrino. If the capture proceeds via $m=0$ transition, the total spin of $^{7}$Li can take values of $F_{f1}=3/2-1/2$ or $F_{f2}=3/2+1/2$. On the other hand, if $m=1$ transition occurs, the total spin can be $F_{f3}=1/2-1/2$ or $F_{f4}=1/2+1/2$. It is evident that both hyperfine levels of $^{7}$Be$^{3+}$ can be contribute to $3/2^{-}\rightarrow3/2^{-}$, but only the upper hyperfine level can contribute to the $3/2^{-}\rightarrow1/2^{-}$ transition. If $^{7}$Be$^{3+}$ exists in the lower hyperfine level, its decay to the first excited state of $^{7}$Li will be blocked, as predicted in Ref.~\cite{Folan1995}. Fig.~\ref{Fig1} shows the decay scheme under hyperfine splitting - solid arrows indicate which transitions are possible. We use the prescription by Patyk~\cite{Patyk2008} to calculate $F(m)$ in $^{7}$Be$^{3+}$ as
\begin{equation}
    \label{Eq7}
    F(m)=
     \begin{cases}
        0, & \text{\textit{m}=1, lower hyperfine level} \\
        8/3, & \text{\textit{m}=1, upper hyperfine level} \\
        1, & \text{\textit{m}=0} \\
     \end{cases}
\end{equation}

\begin{figure}
    \centering
    \includegraphics[width=\linewidth]{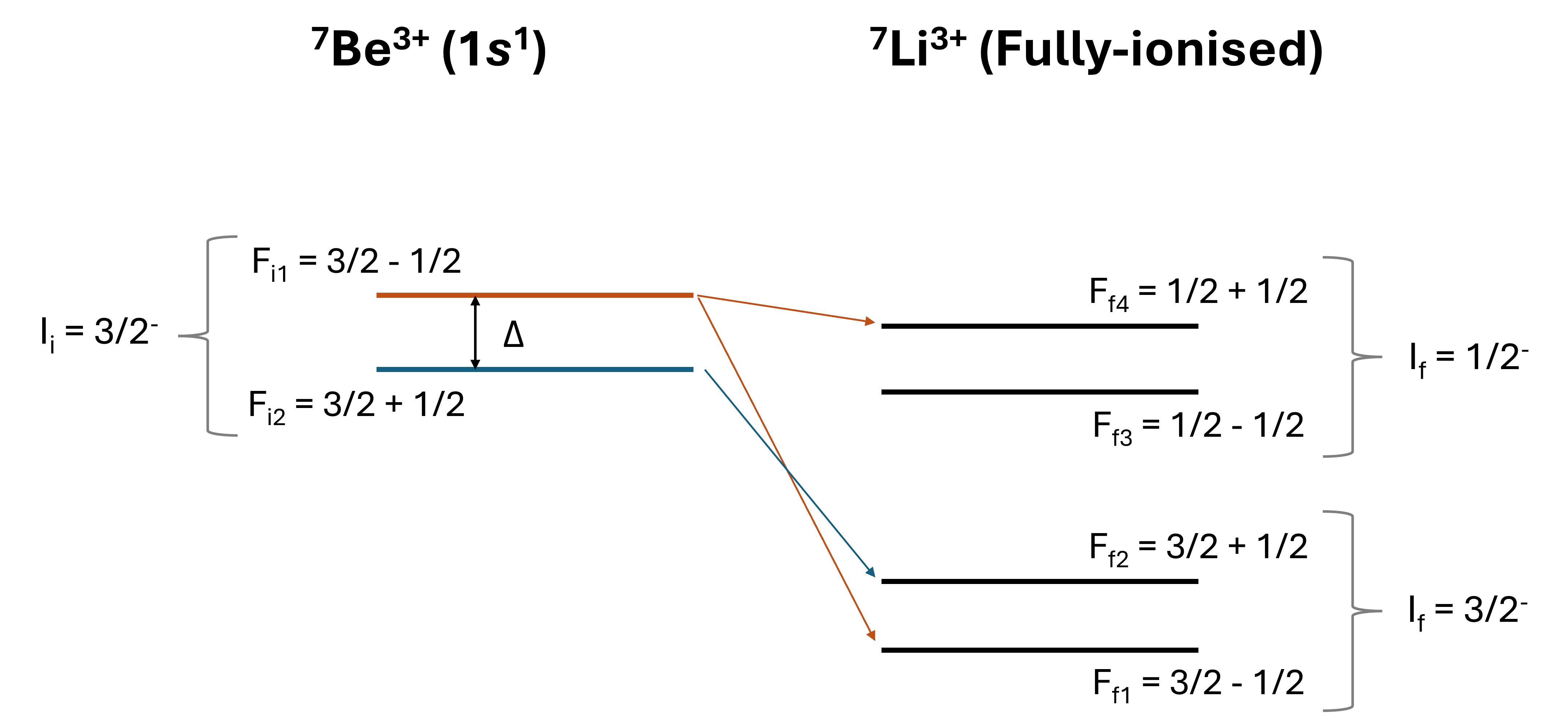}
    \caption{Decay schematic under hyperfine splitting in $^{7}$Be$^{3+}$. No transition from $F_{i2}$ to $F_{f3}$ or $F_{f4}$ is possible.}
    \label{Fig2}
\end{figure}

It can be noted that the transition to the excited state of $^{7}$Li is enhanced by a factor of $8/3$ if the capture proceeds via the upper hyperfine level of $^{7}$Be$^{3+}$, in sharp contrast to its lower energy counterpart. In principle, the transitions from the two hyperfine levels for $m=0$ carry their own weights which must either be experimentally measured or calculated from the matrix element of transition. In the absence of either, we assume $F(0)=1$, implying that the transition proceeds via the base rate with no enhancement/suppression.

\section{Q-Value Calculation}
\label{FLYCHK}

Data on the excitation levels in ions, their ionisation potentials and electronic configurations used in this model are taken from the atomic physics database employed by FLYCHK. FLYCHK is a population kinetics code which can calculate CSD and LPD of ions for different kinds of plasma in various configurations~\cite{FLYCHK2005}. The code is a successor to the FLY suite of codes~\cite{FLY1996} and is capable of generating detailed spectra of plasma in a broad wavelength range. In order to do so, FLYCHK uses an extensive repository of atomic data for elements from $Z=1-79$ which includes valuable info like level electronic configuration, energy, oscillation strength and statistical weight.

The nomenclature of atomic levels in FLYCHK is a combination of existing names from the FLY module~\cite{FLY1996}, the HULLAC database~\cite{HULLAC2001} and superconfiguration states. A detailed explanation of each is provided in Ref.~\cite{FLYCHK2005} but a brief summary is provided here. When performing computations for H-, He- and Li-like ions (meaning one, two and three electrons attached, respectively), FLYCHK uses the FLY database for $Z\leq26$, and the HULLAC database for others. Consequently, levels in $^{7}$Be$^{1/2/3+}$ and $^{7}$Li$^{0/1/2+}$ follow the FLY nomenclature while those in neutral $^{7}$Be$^{0+}$ follow FLYCHK's own system of nomenclature instead. The atomic levels are broadly categorised as bound and autoionising, depending on their energy - the former are levels with energy less than the ionisation energy of the atom/ion whereas the latter are free states that automatically collapse into the next ionisation state. The plasma decay model described in this work only considers bound levels.

Levels in $^{7}$Be$^{0+}$ are named as $04g[n]$ where $[n]$ ranges from $02-10$. The first index $04$ represents $4$ electrons present in the system, while $g$ represents bound states. $[n]$ denotes the principal quantum number of the valence electron but no detailed information on the subshell configuration is available. This is the reason behind the imprecise estimation of $\sigma_{2s}$ in EC rates in neutral $^{7}$Be, which leads to a relative overestimation in the same at $\mathrm{li}2s$.

Levels in $^{7}$Be$^{1+}$ and $^{7}$Li$^{0+}$ are generally designated as $\mathrm{li}[nl]$ where $[nl]$ represents the principal and azimuthal quantum numbers respectively. This detailed notation lasts till $n=5$, after which only $n$ is shown. This implies that subshell configuration details from $\mathrm{li}6$ onward do not exist.

In case of $^{7}$Be$^{2+}$ and $^{7}$Li$^{1+}$, the level notations follow the general structure $\mathrm{he}[nls]$ where $[nls]$ stands for principal and azimuthal quantum numbers of the valence electron, and total spin state respectively. This level of detail holds till $n=2$ beyond which the indices collapse to the simpler form $\mathrm{he}[nps]$.

The levels of $^{7}$Be$^{3+}$ and $^{7}$Li$^{2+}$ are simply represented as the hydrogenic $\mathrm{hy}[n]$ where $n$ is the principal quantum number of the one attached electron. A short summary of the some example level indices and their characteristics is provided in Table~\ref{TabA1}.

As shown in Eq.~\ref{Eq10}, the modification in the Q-value of EC-decay, attributed to the atomic electrons, is the difference in energy between the $ij$- and $i^{\prime}j^{\prime}$-configurations of the parent and daughter ions involved in the decay. For each $^{7}$Be$^{ij}$, the daughter configuration $^{7}$Li$^{i'j'}$ produced depends on the orbital $x$ from which the capture occurs. This is why the superscript in $\Delta\epsilon^{ij,x}$ contains both the charge state/level of $^{7}$Be \textit{and} the orbital $x$. 

To calculate $Q_{m}(ij)$, we first identify the FLYCHK level of $^{7}$Li whose electronic configuration is consistent with the $x$-EC decay of $^{7}$Be and then calculate $\Delta\epsilon^{ij,x}$. The process is repeated for each $i$ and $j$. Tables~\ref{TabEC1}-\ref{TabEC4} explicitly list each level of $^{7}$Be considered in this model and their corresponding $^{7}$Li level generated by capture from either $x=1s_{1/2}$ or $2s_{1/2}$ orbitals. They also list the energy of the levels involved and $\Delta\epsilon^{ij,x}$, which are plotted in Fig.~\ref{Fig3}. 

\begin{table*}
\centering
\caption{Some example FLYCHK level notations and description. A complete list can be found in Ref.~\cite{FLYCHK2005} and its associated manual.}
\label{TabA1}
\begin{tabularx}{\textwidth}{@{\extracolsep{\fill}}ll@{}ll@{}}
\hline
Level & Spectroscopic Notation & Shell Description \\
\hline
$\mathbf{04g02}$  & $(1s)^{2}(2s2p)^{2}$ & Averaged over $l$ ($K^{2}L^{2}$)  \\
$\mathbf{04g07}$  & $(1s)^{2}(2s2p)^{1}(7s7p7d7f7g7h7i)^{1}$ & Averaged over $l$ ($K^{2}L^{1}Q^{1}$)  \\
\hline
$\mathbf{li2s}$  & $(1s)^{2}(2s)^{1}$ & Averaged over $m_{l}$ ($K^{2}L^{1}$)  \\
$\mathbf{li3p}$  & $(1s)^{2}(3p)^{1}$ & Averaged over $m_{l}$ ($K^{2}M^{1}$)  \\
$\mathbf{li7}$  & $(1s)^{2}(7s7p7d7f7g7h7i)^{1}$ & Averaged over $l$ ($K^{2}Q^{1}$)  \\
$\mathbf{op}$ & $(1s)^{1}(2s)^{2}$ & Autoionising level, decays to $\mathbf{li2s}$  \\
\hline
$\mathbf{he1s}$  & $(1s)^{2}$ & Ground state ($K^{2}$)  \\
$\mathbf{he2st}$  & $(1s)^{1}(2s)^{1}\,\,^{3}S$ & Spin triplet ($K^{1}L^{1}$)  \\
$\mathbf{he2ss}$  & $(1s)^{1}(2s)^{1}\,\,^{1}S$ & Spin singlet ($K^{1}L^{1}$)  \\
$\mathbf{he2ps}$  & $(1s)^{1}(2s)^{1}\,\,^{1}S$ & Spin singlet, averaged over $m_{l}$ ($K^{1}L^{1}$)  \\
$\mathbf{2s2s1s}$  & $(2s)^{2}\,\,^{1}S$ & Autoionising level, spin singlet, decays to $\mathbf{he2ps}$ \\   
\hline
$\mathbf{hy1}$  & $(1s)^{1}$ & Ground state ($K^{1}$)  \\
$\mathbf{hy3}$  & $(3s3p3d)^{1}$ & Averaged over $l$ ($M^{1}$)  \\
\hline
\end{tabularx}
\end{table*}

\begin{table*}
\caption{\label{TabEC1}
List of levels in $^{7}$Be$^{0+}$ and the levels of $^{7}$Li$^{0+}$ they can produce.}
\begin{ruledtabular}
\begin{tabular}{c|c|c|c|c|c|c|c}
$^{7}$Be$^{0+,j}$ &
$\mathrm{Energy\,[eV]}$ &
$^{7}$Li$^{0+,j}$ (1s$_{1/2}$) &
$\mathrm{Energy\,[eV]}$ &
$\Delta\epsilon^{ij,1s_{1/2}}\,\mathrm{[eV]}$ & 
$^{7}$Li$^{0+,j}$ (2s$_{1/2}$) &
$\mathrm{Energy\,[eV]}$ &
$\Delta\epsilon^{ij,2s_{1/2}}\,\mathrm{[eV]}$\\
\colrule
$\mathbf{04g02}$ & $0$ & $\mathbf{op}$ & $57.3$ & $-57.3$ & $\mathbf{li2s}$ & $0$ & $0$ \\
$\mathbf{04g03}$ & $7.4$ & $\mathbf{op}$ & $57.3$ & $-49.9$ & $\mathbf{li3s}$ & $4.4$ & $5.6$ \\
$\mathbf{04g04}$ & $8.4$ & $\mathbf{op}$ & $57.3$ & $-48.9$ & $\mathbf{li4s}$ & $4.3$ & $5.6$ \\
$\mathbf{04g05}$ & $8.8$ & $\mathbf{op}$ & $57.3$ & $-49.5$ & $\mathbf{li5s}$ & $4.7$ & $5.6$ \\
$\mathbf{04g06}$ & $8.9$ & $\mathbf{op}$ & $57.3$ & $-48.4$ & $\mathbf{li6}$ & $5.0$ & $5.6$ \\
$\mathbf{04g07}$ & $9.0$ & $\mathbf{op}$ & $57.3$ & $-48.3$ & $\mathbf{li7}$ & $5.1$ & $5.6$ \\
$\mathbf{04g08}$ & $9.1$ & $\mathbf{op}$ & $57.3$ & $-48.2$ & $\mathbf{li8}$ & $5.2$ & $5.6$ \\
$\mathbf{04g09}$ & $9.2$ & $\mathbf{op}$ & $57.3$ & $-48.1$ & $\mathbf{li9}$ & $5.2$ & $5.6$ \\
$\mathbf{04g10}$ & $9.2$ & $\mathbf{op}$ & $57.3$ & $-48.1$ & $\mathbf{li10}$ & $5.3$ & $5.6$ \\
\end{tabular}
\end{ruledtabular}
\end{table*}

\begin{table*}
\caption{\label{TabEC2}
List of levels in $^{7}$Be$^{1+}$ and the levels of $^{7}$Li$^{1+}$ they can produce. $-$ denotes that the corresponding decay is not possible.}
\begin{ruledtabular}
\begin{tabular}{c|c|c|c|c|c|c|c}
$^{7}$Be$^{1+,j}$ &
$\mathrm{Energy\,[eV]}$ &
$^{7}$Li$^{1+,j}$ (1s$_{1/2}$) &
$\mathrm{Energy\,[eV]}$ &
$\Delta\epsilon^{ij,1s_{1/2}}\,\mathrm{[eV]}$ & 
$^{7}$Li$^{1+,j}$ (2s$_{1/2}$) &
$\mathrm{Energy\,[eV]}$ &
$\Delta\epsilon^{ij,2s_{1/2}}\,\mathrm{[eV]}$\\
\colrule
$\mathbf{li2s}$ & $9.3$ & $\mathbf{he2st}$ & $64.4$ & $-55.1$ & $\mathbf{he1s}$ & $5.4$ & $3.9$ \\
$\mathbf{li2p}$ & $13.3$ & $\mathbf{he2pt}$ & $66.7$ & $-53.4$ & - & - & - \\
$\mathbf{li3s}$ & $20.3$ & $\mathbf{he3ps}$ & $75.0$ & $-54.8$ & - & - & - \\
$\mathbf{li3p}$ & $21.3$ & $\mathbf{he3ps}$ & $75.0$ & $-53.8$ & - & - & - \\
$\mathbf{li3d}$ & $21.5$ & $\mathbf{he3ps}$ & $75.0$ & $-53.6$ & - & - & - \\
$\mathbf{li4s}$ & $23.6$ & $\mathbf{he4ps}$ & $77.7$ & $-54.0$ & - & - & - \\
$\mathbf{li4p}$ & $24.0$ & $\mathbf{he4ps}$ & $77.7$ & $-53.6$ & - & - & - \\
$\mathbf{li4d}$ & $24.1$ & $\mathbf{he4ps}$ & $77.7$ & $-53.5$ & - & - & - \\
$\mathbf{li4f}$ & $24.1$ & $\mathbf{he4ps}$ & $77.7$ & $-53.5$ & - & - & - \\
$\mathbf{li5s}$ & $25.1$ & $\mathbf{he5ps}$ & $78.9$ & $-53.8$ & - & - & - \\
$\mathbf{li5p}$ & $25.3$ & $\mathbf{he5ps}$ & $78.9$ & $-53.6$ & - & - & - \\
$\mathbf{li5d}$ & $25.4$ & $\mathbf{he5ps}$ & $78.9$ & $-53.5$ & - & - & - \\
$\mathbf{li5f}$ & $25.4$ & $\mathbf{he5ps}$ & $78.9$ & $-53.5$ & - & - & - \\
$\mathbf{li5g}$ & $25.4$ & $\mathbf{he5ps}$ & $79.5$ & $-53.5$ & - & - & - \\
$\mathbf{li6}$ & $26.0$ & $\mathbf{he6ps}$ & $79.9$ & $-53.5$ & - & - & - \\
$\mathbf{li7}$ & $26.4$ & $\mathbf{he7ps}$ & $80.1$ & $-53.5$ & - & - & -- \\
$\mathbf{li8}$ & $26.7$ & $\mathbf{he8ps}$ & $80.3$ & $-53.4$ & - & - & - \\
$\mathbf{li9}$ & $26.8$ & $\mathbf{he9ps}$ & $80.4$ & $-53.4$ & - & - & - \\
$\mathbf{li10}$ & $27.0$ & $\mathbf{he10p}$ & $80.5$ & $-53.4$ & - & - & - \\
\end{tabular}
\end{ruledtabular}
\end{table*}

\begin{table*}
\caption{\label{TabEC3}
List of levels in $^{7}$Be$^{2+}$ and the levels of $^{7}$Li$^{2+}$ they can produce. $-$ denotes that the corresponding decay is not possible.}
\begin{ruledtabular}
\begin{tabular}{c|c|c|c|c|c|c|c}
$^{7}$Be$^{2+,j}$ &
$\mathrm{Energy\,[eV]}$ &
$^{7}$Li$^{2+,j}$ (1s$_{1/2}$) &
$\mathrm{Energy\,[eV]}$ &
$\Delta\epsilon^{ij,1s_{1/2}}\,\mathrm{[eV]}$ & 
$^{7}$Li$^{2+,j}$ (2s$_{1/2}$) &
$\mathrm{Energy\,[eV]}$ &
$\Delta\epsilon^{ij,2s_{1/2}}\,\mathrm{[eV]}$\\
\colrule
$\mathbf{he1s}$ & $27.5$ & $\mathbf{hy1}$ & $81.0$ & $-53.5$ & - & - & - \\
$\mathbf{he2st}$ & $146.1$ & $\mathbf{hy2}$ & $172.9$ & $-26.7$ & $\mathbf{hy1}$ & $81.0$ & $65.1$ \\
$\mathbf{he2ss}$ & $149.2$ & $\mathbf{hy2}$ & $172.9$ & $-23.7$ & $\mathbf{hy1}$ & $81.0$ & $68.2$ \\
$\mathbf{he2pt}$ & $149.5$ & $\mathbf{hy2}$ & $172.9$ & $-23.4$ & - & - & - \\
$\mathbf{he2ps}$ & $151.2$ & $\mathbf{hy2}$ & $172.9$ & $-21.7$ & - & - & - \\
$\mathbf{he3ps}$ & $167.9$ & $\mathbf{hy3}$ & $189.9$ & $-21.9$ & - & - & - \\
$\mathbf{he4ps}$ & $173.8$ & $\mathbf{hy4}$ & $195.8$ & $-22.0$ & - & - & - \\
$\mathbf{he5ps}$ & $176.5$ & $\mathbf{hy5}$ & $198.6$ & $-22.0$ & - & - & - \\
$\mathbf{he6ps}$ & $178.0$ & $\mathbf{hy6}$ & $200.1$ & $-22.0$ & - & - & - \\
$\mathbf{he7ps}$ & $178.9$ & $\mathbf{hy7}$ & $201.0$ & $-22.0$ & - & - & - \\
$\mathbf{he8ps}$ & $179.0$ & $\mathbf{hy8}$ & $201.6$ & $-22.5$ & - & - & - \\
$\mathbf{he9ps}$ & $179.5$ & $\mathbf{hy9}$ & $202.0$ & $-22.4$ & - & - & - \\
$\mathbf{he10p}$ & $179.9$ & $\mathbf{hy10}$ & $202.2$ & $-22.4$ & - & - & - \\
$\mathbf{he11p}$ & $180.2$ & $\mathbf{hy11}$ & $202.5$ & $-22.3$ & - & - & - \\
$\mathbf{he12p}$ & $180.4$ & $\mathbf{hy12}$ & $202.6$ & $-22.3$ & - & - & - \\
$\mathbf{he13p}$ & $180.5$ & $\mathbf{hy13}$ & $202.7$ & $-22.2$ & - & - & - \\
$\mathbf{he14p}$ & $180.6$ & $\mathbf{hy14}$ & $202.8$ & $-22.2$ & - & - & -- \\
$\mathbf{he15p}$ & $180.7$ & $\mathbf{hy15}$ & $202.9$ & $-22.2$ & - & - & - \\
$\mathbf{he16p}$ & $180.8$ & $\mathbf{hy16}$ & $203.0$ & $-22.2$ & - & - & - \\
$\mathbf{he17p}$ & $180.9$ & $\mathbf{hy17}$ & $203.0$ & $-22.2$ & - & - & - \\
$\mathbf{he18p}$ & $181.0$ & $\mathbf{hy18}$ & $203.1$ & $-22.1$ & - & - & - \\
$\mathbf{he19p}$ & $181.0$ & $\mathbf{hy19}$ & $203.1$ & $-22.1$ & - & - & - \\
$\mathbf{he20p}$ & $181.0$ & $\mathbf{hy20}$ & $203.2$ & $-22.1$ & - & - & - \\
$\mathbf{he21p}$ & $181.1$ & $\mathbf{hy21}$ & $203.2$ & $-22.1$ & - & - & - \\
$\mathbf{he22p}$ & $181.1$ & $\mathbf{hy22}$ & $203.2$ & $-22.1$ & - & - & - \\
\end{tabular}
\end{ruledtabular}
\end{table*}

\begin{table*}
\caption{\label{TabEC4}
List of levels in $^{7}$Be$^{3+}$ and the levels of $^{7}$Li$^{3+}$ they can produce. $-$ denotes that the corresponding decay is not possible.}
\begin{ruledtabular}
\begin{tabular}{c|c|c|c|c|c|c|c}
$^{7}$Be$^{3+,j}$ &
$\mathrm{Energy\,[eV]}$ &
$^{7}$Li$^{3+,j}$ (1s$_{1/2}$) &
$\mathrm{Energy\,[eV]}$ &
$\Delta\epsilon^{ij,1s_{1/2}}\,\mathrm{[eV]}$ & 
$^{7}$Li$^{3+,j}$ (2s$_{1/2}$) &
$\mathrm{Energy\,[eV]}$ &
$\Delta\epsilon^{ij,2s_{1/2}}\,\mathrm{[eV]}$\\
\colrule
$\mathbf{hy1}$ & $181.4$ & Fully-ionised & $203.5$ & $-22.0$ & - & - & - \\
$\mathbf{hy2}$ & $344.5$ & - & - & - & Fully-ionised & $203.5$ & $141.0$ \\
\end{tabular}
\end{ruledtabular}
\end{table*}

\end{document}